\documentclass[11pt,a4paper,onecolumn,eqsecnum,eprintnumbers,floatfix,aps,prd,nofootinbib,amsmath,amssymb,amsfonts,superscriptaddress,twoside]{revtex4-2}
\usepackage{graphicx,rotating,hyperref,slashed,amsmath,xcolor,colortbl}
\usepackage{tabularx}
\usepackage{empheq}
\usepackage[utf8]{inputenc} 
\usepackage[T1]{fontenc}

\usepackage[normalem]{ulem}
\usepackage{tikz}
\usepackage{mciteplus} 
\usepackage{enumitem}
\usepackage{comment}
\usepackage{booktabs}

\makeatletter
\allowdisplaybreaks	 

\def\rotangle{\psi}
\def\beq{\begin{equation}}
\def\eeq{\end{equation}}

\newcommand{\bv}{{\bm v}}
\newcommand{\bn}{{\bm n}}

\newcommand{\Ds}{{\cal D}_s}
\newcommand{\Do}{{\cal D}_o}

\newcommand*\widefbox[1]{\fbox{\hspace{2em}#1\hspace{2em}}}

\renewcommand\[{\left[}

\def\bea{\begin{eqnarray}}
\def\eea{\end{eqnarray}}

\def\dd{{\rm d}}
\def\ii{{\rm i}}

\def\Bsource{{\cal B}}

\newcommand{\be}{\begin{equation}}
\newcommand{\ee}{\end{equation}}

\def \bm#1{\mbox{\boldmath$#1$\unboldmath}} 

\definecolor{rossos}{cmyk}{0,1,1,0.55}
\definecolor{blu}{cmyk}{1,1,0,0.3}
\definecolor{bluc}{cmyk}{1,1,0,0.1}
\definecolor{verde}{cmyk}{0.92,0,0.59,0.25}
\definecolor{verdec}{cmyk}{0.92,0,0.59,0.15}
\definecolor{verdes}{cmyk}{0.92,0,0.59,0.4}
\hypersetup{colorlinks, bookmarksopen, bookmarksnumbered,
citecolor=blu, linkcolor=bluc, pdfstartview=FitH, urlcolor=rossos}

\definecolor{mygreen}{rgb}{0,0.6,0}

\begin{document}

\title{Boosting gravitational waves: a review of kinematic effects on amplitude, polarization, frequency and energy density}

\author{Giulia Cusin}
\email{cusin@iap.fr}
\affiliation{Institut d'Astrophysique de Paris, UMR-7095 du CNRS et de Sorbonne Universit\'e, Paris, France}
\affiliation{Département de Physique Théorique and Center for Astroparticle Physics, Université de Genève, Quai E. Ansermet 24, CH-1211 Genève 4, Switzerland}
\author{Cyril Pitrou}
\email{pitrou@iap.fr, corresponding author}
\affiliation{Institut d'Astrophysique de Paris, UMR-7095 du CNRS et de Sorbonne Universit\'e, Paris, France}
\author{Camille Bonvin}
\email{camille.bonvin@unige.ch}
\affiliation{Département de Physique Théorique and Center for Astroparticle Physics, Université de Genève, Quai E. Ansermet 24, CH-1211 Genève 4, Switzerland}
\author{Aur\'elien Barrau}
\email{barrau@lpsc.in2p3.fr}
\affiliation{Laboratoire de Physique Subatomique et de Cosmologie, Univ. Grenoble-Alpes, CNRS-IN2P3, 53 av. des Martyrs, 38026 Grenoble, France}
\author{Killian Martineau}
\email{martineau@lpsc.in2p3.fr}
\affiliation{Laboratoire de Physique Subatomique et de Cosmologie, Univ. Grenoble-Alpes, CNRS-IN2P3, 53 av. des Martyrs, 38026 Grenoble, France}

\date{\today}

\begin{abstract}
\noindent
We review the kinematic effects on a gravitational wave due to either a peculiar motion of the astrophysical source emitting it or a local motion of the observer. Working in the context of general relativity, we show at fully non-linear order in velocity, that the amplitude of the wave is amplified by the Doppler factor in the case in which the source moves with respect to a reference frame, while it is invariant if the observer moves (with respect to a reference observer). However, the observed specific intensity transforms in the same way under a boost of the source or of the observer. We also show at fully non-linear order that under a boost (of either source or observer), the polarization tensor is rotated in the same way the wave direction is rotated by aberration, such that the only net effect of a boost on polarization is to change the phase of the helicity components. We apply these results to a wave emitted by a binary system of compact objects in the cosmological context. 
\end{abstract}
\maketitle

\section{Introduction}

Gravitational waves (GWs) provide a fantastic new window onto our Universe. In particular, the form of the GW depends intimately on the properties of the sources, allowing us to probe their nature; and it also contains information about the theory of gravity, allowing us to test the validity of General Relativity in various regimes. GW waveforms are usually derived assuming that the source and the observer are at rest with respect to the Hubble flow, i.e.\ that their relative peculiar velocity is zero. In practice however, this is not an accurate assumption, since GW sources are affected by the structures in our Universe. They move inside galaxies, at velocities that can be quite significant, and the galaxies themeselves move with respect to the Hubble flow. Over the past years, several works have studied the impact of kinematic effects on GW observables in the cosmological context. In particular, Refs.~\cite{Bonvin:2016qxr, Inayoshi:2017hgw, Tamanini:2019usx,Sberna:2022qbn} studied the effect of the source peculiar velocity and acceleration on the chirp signal from a binary system of compact objects, while Ref.~\cite{Bonvin:2022mkw} derived the effect of aberration of the wave polarization at linear order in the velocity. In parallel, Refs.~\cite{Cusin:2022cbb, Tasinato:2023zcg} studied kinematic effects on a stochastic background of gravitational waves due to a peculiar motion of the observer with respect to the source emission frame. The impact of a peculiar motion of the observer on a distribution of GW sources was studied in \cite{Mastrogiovanni:2022nya, Kashyap:2022ibx,Stiskalek:2020wbj,Essick:2022slj,Grimm:2023tfl}.

The goal of this article is to provide a complete review and first principles derivation of all possible kinematic effects  on gravitational wave observables. Our analysis is valid in the context of general relativity and on a generic background, and at fully non-linear order in the velocity. We clarify some subtle points 
related to the different role of observer and source motion, that have been previously debated in the literature~\cite{1999PThPh.101..903S, Bonvin:2005ps, Hui:2005nm, Kaiser:2014jca}.  We also stress analogies and differences of the gravitational wave case with respect to the electromagnetic counterpart. 

In our derivation, we compare two separate situations, that we show to be physically not equivalent: A) a source moving with respect to a reference source located at the same position when the signal is \emph{emitted}; and B) an observer (detector) moving with respect to a reference observer located at the same position when the signal is \emph{received}. 

Using the eikonal approximation to study GW propagation, we identify the following three effects of a boost:
\begin{enumerate}
    \item {\bf{amplitude}}: it is amplified by the Doppler factor in case A, where the source moves with respect to a reference source, while it is invariant in case B, where the observer moves with respect to a reference observer; 
    
    \item {\bf{polarization tensor}}: we explicitly  show at fully non-linear order that the polarization tensor is rotated in the same way the wave direction is rotated by aberration, such that the only net effect of a boost on polarization is to change the phase of the helicity components. This transformation occurs in both cases A and B, i.e.\ if the source moves with respect to a reference source, or if the observer moves with respect to a reference observer, but the intrinsic properties of the measured polarization are only modified in case A.\footnote{We will detail that in case A, as a consequence of aberration, we do not receive the same signal (we see another side of the source, hence the degree of polarisation is not the same), while in case B, the signal received is the same, but the source appears at an aberrated direction.}

    \item {\bf{frequency}}: it is enhanced by the Doppler factor, also in both cases A and B. More precisely, there is a blueshift if the source moves with respect to a reference source in the direction of the observer, or if the observer moves with respect to a reference observer in the direction of the source.

    \item {\bf{energy density}}: it is enhanced by four powers of the Doppler factor in case A, whereas it is only enhanced by two powers of the Doppler factor in case B. However, the total specific intensity, which is the flux of energy propagating in a given solid angle, transforms symmetrically with four powers of the Doppler factor in both cases.
    
\end{enumerate}

We stress that the fact that the source velocity affects the strain amplitude of the GW, while the observer velocity does not, is a general property, valid in a generic space time (also in Minkowski). This is due to the fact that the wave amplitude scales as $1/\chi$ where $\chi=\sqrt{\dd A_o/\dd\Omega_s}$, where $\dd\Omega_s$ corresponds to the angle of the emitted bundle that is received by the observer and $\dd A_o$ is the surface of the bundle at the observer. This quantity is the same for all observers related by a boost (at the observer position) but it is not the same if we have two sources moving with different velocities at the same point of emission. This apparent asymmetry between the motion of a source and the motion of an observer does not break the equivalence principle but it is simply related to the fact that cases A and B are two different physical situations, in which the distances between the source and the observer are different, as detailed in section~\ref{boostwave}.

As a consequence of these transformation rules, we explicitly show that for a given source there exists a whole class of equivalent sources, related by boosts and rotations, that give rise to exactly the same observed signal, i.e.\ same amplitude, frequency and polarization tensor. This implies that the source velocity cannot be inferred from an analysis of the received signal.\footnote{In contrast, if the source velocity varies with time, the acceleration of the source can be measured from the waveform~\cite{Bonvin:2016qxr, Inayoshi:2017hgw, Tamanini:2019usx}.}

The rest of the paper is structured as follows: in section \ref{eikonal} we review the basic equations governing the propagation of a  GW in the eikonal approximation. We show that the GW amplitude scales as the inverse of the distance $\chi$, whose transformation under boosts of either the source or the observer can be easily inferred. In section \ref{boost} we review the basic transformation rules between observables defined in two frames related by a boost. This is applied in section \ref{boostwave} to infer the transformation properties of the GW amplitude, energy density, polarization tensor and frequency.  The observational implications of these transformation rules are discussed in section~\ref{SecObsImplications}, where we build the family of equivalent sources generating the same observed signal. In section \ref{cosmo} we turn to a cosmological setting and we apply these general results to illustrate how a GW emitted by a binary system of compact objects in the Newtonian approximation is affected by peculiar motions of the source and the observer (with respect to the Hubble flow reference frame). Finally in section \ref{conclusions} we summarize the key points of our article, and in a series of appendices we explicit the analogy with the electromagnetic case and geometric optics.

We use $c=1$ (we put it back only where clarity requires it), and also $\hbar=1$ while we keep track of the gravitational constant $G$. 

\section{Propagation of GWS in the eikonal approximation}\label{eikonal}

\subsection{WKB expansion}\label{SecWKB}

Let us consider the eikonal approximation to study GW propagation (see e.g. \cite{Isaacson:1967zz}, \cite{Cusin:2019rmt} and chapter 1 of \cite{Maggiore:1900zz}). We write the metric as 
\begin{equation}\label{ansatz}
	h_{\mu\nu}= \Re(H_{\mu\nu}e^{i\omega \Phi})\,,
\end{equation}
where $\Re$ is the real part of the expression in parenthesis. We assume that the phase $\Phi$ varies much faster than the wave amplitude $H_{\mu\nu}$, hence we introduce the large dimensionless parameter $\omega$ that governs a WKB type expansion. 

At leading and next-to-leading order in the geometric optics parameter $\omega$, Einstein equations\footnote{When treating $h_{\mu\nu}$ as a traceless perturbation around a background with metric $g_{\mu\nu}$, with Levi-Civita connection $\nabla_\mu$ and Riemann tensor $R_{\alpha\beta\mu\nu}$, the linearized Einstein equation leads to $\nabla^\alpha \nabla_\alpha h_{\mu\nu} + 2R_{\alpha\mu\beta\nu} h^{\alpha\beta} - 2 R_{\alpha (\mu} h^\alpha_{\,\,\nu)} =0$ if the harmonic gauge condition $\nabla^\mu h_{\mu\nu}=0$ is satisfied. The term involving the Ricci and the Riemann tensor can be discarded because they are of order ${\cal O}(\omega^0)$ in the WKB expansion. The GW wave equation eventually reduces to $\nabla^\mu \nabla_\mu h_{\alpha\beta}=0$, which with the ansatz~\eqref{ansatz} leads to \eqref{geomopticsexpansion}.} lead to the following equation that describes the geometric optics regime:
\begin{equation}\label{geomopticsexpansion}
-\omega^2 k_\beta k^\beta H_{\mu\nu}+i\omega [2 k^\beta \nabla_{\beta}H_{\mu\nu}{}+\nabla_{\beta}k^\beta H_{\mu\nu}]+\mathcal{O}(\omega^0)=0\,,
\end{equation}
where the eikonal momentum is $k_{\mu}=\partial_{\mu}\Phi$.
This leads to the two conditions for each order in $\omega$:
\begin{subequations}
\begin{align}
& k_\beta k^\beta=0\,,\label{go:null}\\
&2 k^\beta \nabla_{\beta}H_{\mu\nu}{}+(\nabla_{\beta}k^\beta) H_{\mu\nu}=0\,.\label{go:linear}
\end{align}
\end{subequations}
From eq.~(\ref{go:null}) we see that $k^\mu$ is a null vector. Since eq.~(\ref{go:null}) implies $k^\mu \nabla_\nu k_\mu =0$, and with $\partial_\mu \partial_\nu \Phi = \partial_\nu \partial_\mu \Phi$, this leads directly to the null geodesic equation
\begin{equation}\label{geodesic}
k^\mu \nabla_{\mu}k_{\nu}=0\,.
\end{equation}
Therefore, in the eikonal approximation, we can define trajectories for gravitational waves which are null geodesics, $x^\mu(\lambda)$, with tangent vector $\dd x^\mu /\dd \lambda = k^\mu$ always normal to the constant $\Phi$ null hypersurfaces. We also deduce that the phase is conserved along the geodesics: $k^\mu k_\mu = (\dd x^\mu/\dd \lambda) \partial_\mu \Phi = \dd \Phi/\dd \lambda = 0$.  Moreover, the Lorenz gauge condition $\nabla^{\mu}h_{\mu\nu}=0$ at leading order in $\omega$ leads to 
\begin{equation}
k^\mu H_{\mu\nu}=0\,,
\end{equation}
which indicates that $H_{\mu\nu}$ is a transverse tensor.\footnote{An estimate of the regime in which beyond-WKB corrections become relevant can be found in \cite{Cusin:2018avf} around eq.~(2.4): beyond-WKB corrections become relevant for $\lambda > r^{3/2}/M^{1/2}>b^{3/2}/M^{1/2}$, where $b$ is the impact parameter to the lens. 
This means that for a fixed finite wavelength, corrections to geometric optics will appear only in situation of special alignment with compact structures along the line of sight. This probability is estimated in \cite{Cusin:2018avf} to be very small in general. However, for special astrophysical systems such as triple systems studied in \cite{Pijnenburg:2024btj}, beyond geometric optics effect are important for a large range of configurations.}

\subsection{Amplitude and polarization}
\label{sec:amplitude_gen}

We split $H_{\mu\nu}$ into an amplitude and a polarization unit tensor as
 \be\label{AAA}
 H_{\mu\nu}=H  \epsilon_{\mu\nu}\,,
 \ee
 with 
\begin{subequations}
\begin{align}
 &H = \sqrt{H^{*}_{\mu\nu}H^{\mu\nu}}\,,\qquad\epsilon_{\mu\nu}\epsilon^{\mu\nu}=1\,,\\ &k^\mu \epsilon_{\mu\nu} = 0\,,\label{Gaugeepsilon}
 \end{align}
\end{subequations}
 which allows us to recast eq.~(\ref{go:linear}) as two separate equations
\begin{subequations}\label{conservationall}
\begin{align}
& k^\mu \nabla_{\mu}H=-\frac{1}{2}\theta H; \quad \theta = \nabla_{\mu}k^\mu\, ,\label{conservation}\\
& k^\alpha \nabla_{\alpha} \epsilon_{\mu\nu}=0 \label{conservation2}\,.
\end{align}
\end{subequations}
The second equation indicates that polarization is parallel-propagated along the null vector $k^\mu$, while the first one leads to the covariant conservation of the flux (i.e.~$\nabla_\alpha(H^2k^\alpha)=0$), which can be interpreted as the conservation of the number of gravitons. In Eq.\,(\ref{conservation}), 
$\theta$ is the Sachs scalar which characterizes the divergence of the geodesic beam.  Eq.\,(\ref{conservation}) can be solved using that the Sachs scalar is related to the variation of distances (valid at fully non linear order, see eq.~(2.85) of \cite{Fleury:2015hgz}) by
\be
\theta=\nabla_{\alpha}k^{\alpha}=2 \frac{\dd\log \chi}{\dd\lambda}\,.
\ee
Here $\chi$ is a distance defined as
\be\label{chi}
\chi=\sqrt{\frac{\dd A_o}{\dd\Omega_s}}\,,
\ee
where $\dd\Omega_s$ corresponds to the solid angle of the emitted bundle of geodesics that is received by the observer and $\dd A_o$ is the surface of the bundle at the observer.\footnote{Note that the surface and the shape of a bundle (where it is not singular) does not depend on the observer measuring it at a specific location. See section~\ref{secdistances} for a proof.} 
From now on we denote with a subscript $s$ quantities at the source and with a subscript $o$ quantities at the observer. We obtain
\be\label{ampl}
H(\lambda)=H(\lambda_s)\frac{\chi(\lambda_s)}{\chi(\lambda)}\,,
\ee
where $\lambda_s$ has to be interpreted as some affine parameter close to the source. The GW amplitude therefore scales as $1/\chi=\sqrt{\dd\Omega_s/\dd A_o}$, and the quantity
\be\label{DefB}
\Bsource \equiv H(\lambda_s) \chi(\lambda_s)
\ee
is an intrinsic property of the signal emitted  by the source which does not depend on the $\lambda_s$ used to define it. 

Equations~\eqref{conservationall} are supplemented with corrections when beyond geometric optics corrections are taken into account~\cite{Andersson:2020gsj, Cusin:2019rmt}. Since the definition~\eqref{chi} of the distance $\chi$ relies on the definition of null rays (bundles), then if eq.~\eqref{conservation} does not hold in a modified gravity theory, i.e. if GWs do not follow null rays (e.g. in massive gravity), one has to introduce a different definition for the distance $\chi$, hence all our results need to be revised. In case in which the dispersion relation~\eqref{conservation} still holds, one could still find modifications to~\eqref{conservation2}, hence find different behaviours for polarisation transport and/or for the scaling of the wave amplitude with distance.

\begin{figure}[!ht]
\centering
\includegraphics[width=0.87\textwidth]{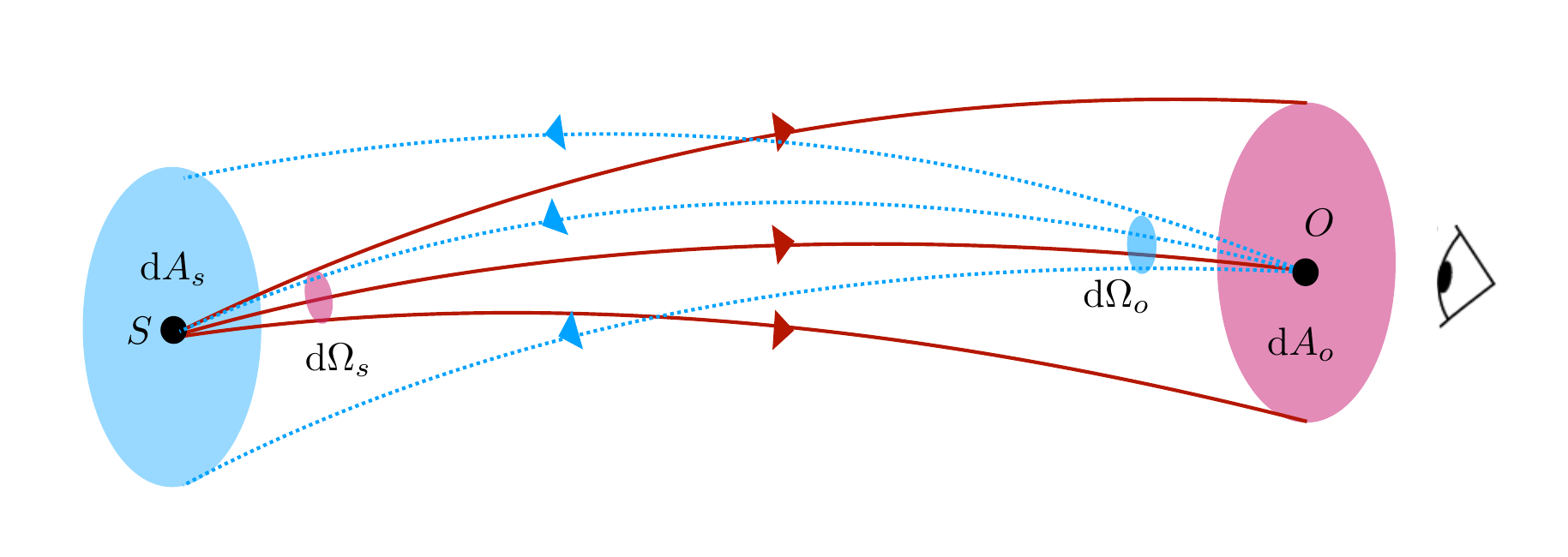}
\caption{Schematic representation of the beam from the source reaching the observer (in red), and of the fictitious beam sent from the observer backward in time to the source (in blue). }
\label{FigDistances}
\end{figure}

As illustrated in Fig.~\ref{FigDistances}, one can define another notion of distance, just by interchanging the role of source and observer. This gives rise to the angular diameter distance
\be
D_A=\sqrt{\frac{\dd A_{s}}{\dd\Omega_o}}\,,
\ee
which relates the solid angle at the observer to the surface at the source. In Fig.~\ref{FigDistances}, we show  that the exchange of the role of source and observer, can be interpreted as if the observer was irradiating the source with a beam whose opening angle is $\dd\Omega_o$, going backward in time.

 These distances are usually introduced in the electromagnetic context and they require the concept of rays. It follows that they can be extended also to the gravitational wave context, as long as the geometric optics approximation is valid (even if rays are not observable in that context). These two distances are related by the reciprocity relation (see e.g. eq.~(3.67) in \cite{1992grle.bookS} or eq.~(3.30) of~\cite{Fleury:2015hgz})
\be\label{Eq.chitoDA}
\chi=(1+z)D_A\,,
\ee
where the source redshift is defined as 
\be\label{z}
(1+z)=\frac{u^{\mu}_{s}k_{\mu}}{u_o^{\mu}k_{\mu}}\,,
\ee
with $u^{\mu}_{s}$ and $u_o^{\mu}$ the source and observer 4-velocities.

We observe that the distance $\chi$ is the distance measured by the source to the observer \emph{when the signal is received}. On the other hand, the angular diameter distance is the distance $\chi$ that is obtained by interchanging the role of source and observer: the observer sends a bundle back in time to irradiate the source. 
It corresponds to a distance measured by the observer to the source \emph{at emission time}. 

\subsection{Energy and energy density}

For future use, it is useful to derive the energy momentum tensor associated with the GW, in this geometric optics picture. 
We specialise the covariant expression of the energy momentum tensor (see e.g. eq. (1.125) of \cite{Maggiore:1900zz}) to a locally inertial frame, at the observer position. One can then verify that the pseudo energy momentum tensor at leading order in geometric optics is equivalent to the result on Minkowski\footnote{The expression on a flat background and the covariant one (specialised to a local inertial frame) differ by terms $\propto\partial \Gamma h$. The derivative of the Christoffel on a locally inertial frame is related to the local expression of the background Riemann tensor, hence in geometric optics the difference is negligible in front of the leading terms $\propto (\partial h)^2$.}
\be
t_{\mu\nu}=\frac{c^4}{32 \pi G}\left[ \partial_{\mu} h_{\alpha \beta}\partial_{\nu}h^{\alpha\beta}\right]_{\rm av}\,,
\ee
where the square brackets denote an average over several periods of the wave. At leading order in geometric optics we obtain
\be\label{tmunuGW}
t_{\mu\nu}=\frac{c^4}{64 \pi G} H^2 k_{\mu}k_{\nu}\,,
\ee
and the associated energy density for a given observer with velocity $u^\mu=\dd x^\mu/\dd \tau$ is
\be\label{rho}
\rho_{\rm GW}=\frac{c^2}{64 \pi G} H^2 \left(k_{\mu}u^{\mu}\right)^2\,.
\ee

The scalar quantity $-\left(k_{\mu}u^{\nu}\right)$ is interpreted as the energy $E$ measured by the observer. It is associated with the rate of change of the phase seen by an observer with velocity $u^{\mu}$ since
\be\label{dPhidtau}
-\frac{\dd\Phi}{\dd\tau} = -u^{\mu}\partial_{\mu}\Phi = -u^{\mu}k_{\mu} = E = 2\pi f\,,
\ee
where $f$ is the GW frequency.

From eq.~\eqref{rho} we introduce the ``luminosity distance'' $D_L$, defined such that the GW energy density scales as $1/D^2_L$. From the scaling~\eqref{ampl} and the definition of redshift~\eqref{z}, this latter distance is related to the other distances by
\be\label{DefDL}
D_L = (1+z)\chi = (1+z)^2 D_A\,.
\ee
From this relation, we see that the luminosity distance $D_L$ that governs the scaling of the GW energy density is nothing else than the standard luminosity distance defined for electromagnetic sources~\cite{1992grle.bookS}. 
With this we can write eq.~\eqref{ampl} in the alternative forms 
\be\label{amplitude}
H(\lambda) = \frac{\Bsource}{(1+z(\lambda)) D_A(\lambda)} = \Bsource \frac{(1+z(\lambda))}{D_L(\lambda)} \,.
\ee
Hence, whatever the waveform considered, the GW amplitude (hence the strain) scales as $1/\chi=1/[(1+z)D_A]=(1+z)/D_L$.

Note that all the results of this section can also be obtained in the case of electromagnetism and light propagation along rays in the geometric optics approximation. One needs only to consider an expansion for the potential vector $a_\mu$ analogous to the expansion~\eqref{ansatz}, see appendix~\ref{SecEikonalA} for the essential results.

\section{Transformations under a boost}\label{boost}

Since we are interested in computing the impact of the source and observer velocity on GWs, we start by reviewing the transformation properties of observable quantities defined in two frames related by a boost. 

\subsection{Boost components}

We associate to each observer a frame described by a tetrad, that is an orthonormal basis of four vectors. They are noted $e_a^\mu=(u^\mu,e_i^\mu)$ and $\tilde e_a^\mu =(\tilde u^\mu,{\tilde e_i}^\mu)$. The first vector of each tetrad is the observer 4-velocity, and the remaining three are a set of orthonormal spatial vectors. These tetrads are related by~\cite{Cusin:2016kqx} 
\be\label{DefBoost1}
\tilde e_a^\mu = e_b^\mu
{({\Lambda^{-1}})^b}_a = {\Lambda_a}^b e_b^\mu\,,
\ee
where 
\be
{\Lambda_0}^0 = \gamma \,,\,\quad {\Lambda_0}^i = {\Lambda_i}^0 = 
\gamma v_i\,,\qquad {\Lambda_i}^j = \delta^j_i +
\frac{\gamma^2}{1+\gamma}v_i v^j\,,
\ee
with $\gamma\equiv 1/\sqrt{1-\beta^2}$ and $\beta^2 \equiv v_i v^i$. In particular the 4-velocities in the two frames are related covariantly by  
\be
{\bm{\tilde{u}}} = \gamma({\bm u} + {\bm v}),\qquad {\bm{{u}}} = \gamma({\bm{\tilde{u}}} - {\bm{\tilde{v}}})\,,
\ee
where the spatial boost velocities defined by each observer are
\be\label{DefBoostVelocity}
{\bm v} \equiv v^i {\bm e}_i,\qquad {\bm{\tilde{v}}} \equiv v^i {\bm{\tilde{e}}}_i\,.
\ee

\subsection{Doppler effect}

We decompose the eikonal momentum $k^\mu$ in two different ways, depending on the frame in which it is observed, as
\be\label{kdecomposition}
{\bm k} = E ({\bm u} - {\bm n}) = \tilde{E}({\bm{\tilde{u}}}- {\bm{\tilde{n}}})\,,\qquad\qquad {\bm u} \cdot {\bm n} = {\bm{\tilde{u}}} \cdot {\bm{\tilde{n}}} =0\,.
\ee
Note that $k^\mu$ is an intrinsic property of the wave, which depends on the spacetime geometry but not on the observer. It is therefore the same in both frames. Its projection on the observer tetrad is, on the other hand, observer-dependent.
With this choice ${\bm n}$ and $\bm{\tilde{n}}$ correspond to the directions of observation, whereas the directions of propagation of the wave are $-{\bm n}$ and $-\bm{\tilde{n}}$. 
The energies or frequencies in both frames are related by 
\be\label{TRuleE}
\tilde E = - {\bm{\tilde{u}}} \cdot {\bm k}  \equiv \mathcal{D} E\,,
\ee
where the Doppler shift factor is defined as
\be\label{alphadef}
\mathcal{D} \equiv  \gamma (1+ {\bm n}\cdot
{\bm v})=\frac{1}{\gamma(1-{\bm {\tilde{n}}}\cdot {\bm{\tilde{v}}})}\,.
\ee
We define the components of the direction vectors in their respective frame by 
\be\label{DefComponents}
n_i \equiv {\bm n} \cdot {\bm e}_i\,,\qquad \widetilde{n_i} \equiv \bm{\tilde n} \cdot \bm{\tilde e}_i\,,
\ee
where in the last notation ($\widetilde{n_i}$), the tilde stretches over both the vector and the components to emphasize that not only it is not the same vector, but its components are not read in the same basis.
With these definitions, the scalar products in \eqref{alphadef} are expressed with definitions \eqref{DefBoostVelocity} as ${\bm n}\cdot {\bm v} = n_i v^i$ and ${\bm {\tilde{n}}}\cdot {\bm{\tilde{v}}} = \widetilde{n_i} v^i$. 

\subsection{Aberration}

From the decomposition \eqref{kdecomposition} and the Doppler shift definition~\eqref{alphadef}, the aberrated direction is simply given covariantly by
\be
{\bm{\tilde{n}}} = {\bm{\tilde{u}}} + \mathcal{D}^{-1}({\bm n} - {\bm u})\,.
\ee
However in practical situations we are interested in relating the components~\eqref{DefComponents}, and from the transformation of the basis~\eqref{DefBoost1} we find $\widetilde{n_i}=\mathcal{D}^{-1}\left(\Lambda_i^{\,\,j} n_j + \Lambda_i^{\,\,0}\right)$, that is 
\be\label{Eq:ntilde1}
\widetilde{n_i}= \frac{1}{\mathcal{D}}\left(n_i
  +\frac{\gamma^2}{1+\gamma} {\bm n} \cdot {\bm v} v_i + \gamma v_i\right)\,. 
\ee
Defining ${\bm {\hat{v}}}$ as the unit vector in direction of ${\bm v}$ such that ${\bm v} = \beta {\bm {\hat{v}}}$ and $v^i = \beta \hat{v}^i$, the expression for aberration is more often rewritten in terms of components along and orthogonally to $v^i$ as 
\be\label{Eq:ntilde2}
\widetilde{n_i} = \frac{\hat v_i}{1+{\bm n}\cdot{\bm v}}\left(n_j \hat
  v^j + \beta\right)+\frac{1}{\gamma(1+{\bm n}\cdot{\bm v})}(n_i-{\bm n}\cdot \bm{\hat{v}} \hat{v}_i) \,, 
\ee
which leads to the usual aberration formula
\be\label{aberrnv}
\widetilde{\bm n}\cdot {\bm{\hat{\tilde{v}}}}= \widetilde{n_i} \hat{v}^i =\frac{{\bm n}\cdot {\bm{ \hat v}}+\beta}{1+{\bm n}\cdot {\bm v}} \,.
\ee
Notice that the aberration equation \eqref{Eq:ntilde1} can also be reformulated in compact form as 
\be\label{aberr}
\widetilde{n_i}= \frac{1}{\mathcal{D}}\left(n_i + \alpha \hat{v}_i\right)\,,\qquad \mbox{with}\quad \alpha \equiv (\gamma-1){\bm n} \cdot {\bm {\hat{v}}} + \gamma \beta\, ,
\ee
and the normalisation of $\widetilde{n_i}$ implies the identity
\be\label{id}
1+ 2 \alpha {\bm n} \cdot {\bm {\hat{v}}} + \alpha^2 = \mathcal{D}^2\,.
\ee
Finally, in appendix \ref{Cyril} we show that aberration can be recast as 
\be\label{ntildeRni}
\widetilde{n_i} = R_i^{\,\,j} n_j\,, 
\ee
where $R_i^{\,\,j}\equiv R_i^{\,\,j}(v_k,n_k)$ is a rotation whose characteristics depend on ${\bm n}$ and ${\bm v}$, which is defined by~\eqref{JijtoRij}, and whose explicit expression is given by~\eqref{Rijresummed}. 

\subsection{Adapted coordinates}\label{SecAdapted}

\begin{figure}[!htb]
\includegraphics[width=0.49\textwidth]{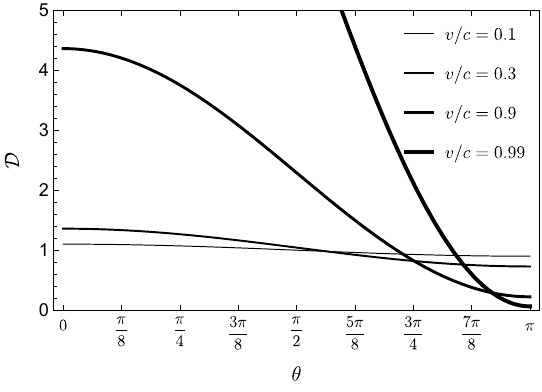}
\includegraphics[width=0.49\textwidth]{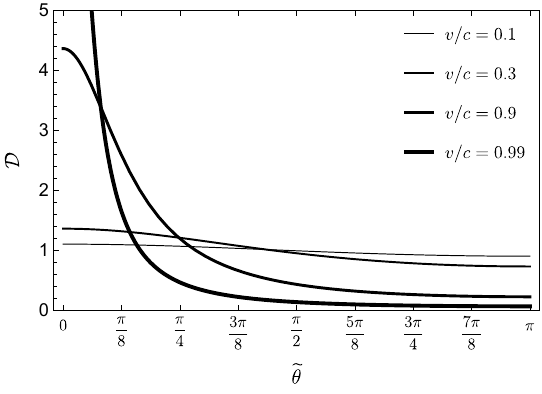}\\
\includegraphics[width=0.49\textwidth]{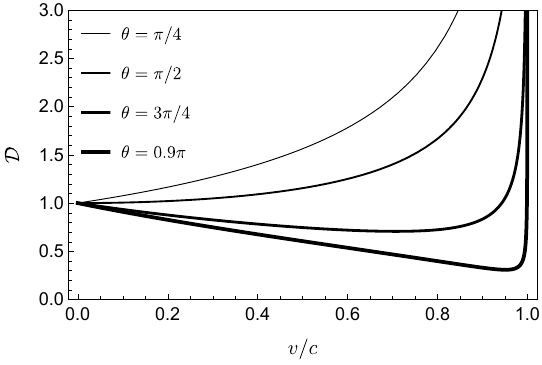}
\includegraphics[width=0.49\textwidth]{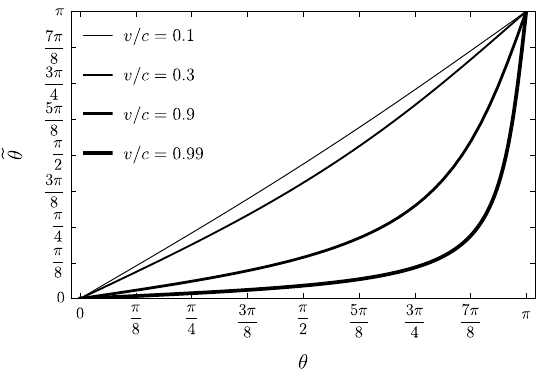}
\caption{{\it Top left:} Doppler shift factor ${\cal D}$ as a function of the initial direction angle (before the boost) $\theta$ for several values of $\beta = v/c$. {\it Top right:} ${\cal D}$ as function of the final direction angle (after the boost) $\tilde \theta$.  {\it Bottom left:} ${\cal D}$ as function of $\beta$ for several initial directions. For $\theta > \pi/2$, ${\cal D}$ first decreases, then reaches a minimum when $\beta = - \cos \theta$, and subsequently increases until diverging when $\beta \to 1$. {\it Bottom right:} Aberrated angle $\tilde \theta$ as a function of the initial direction angle $\theta$. All directions (except $\theta=\pi$) are attracted toward $\tilde \theta = 0$ when $\beta \to 1$. See also Fig.~\ref{FigAberration} for an illustration of the aberration of directions and the effect on polarization.}
\label{DopAngle}
\end{figure}

When the spherical coordinates associated to the zenith direction ${\bm{ \hat v}}$ are chosen, that is ${\bm{ \hat v}} = {\bm e}_z$ and ${\bm n} = \sin \theta(\cos \phi {\bm e}_x + \sin \phi {\bm e}_y) + \cos \theta {\bm e}_z$, the Doppler shift factor \eqref{alphadef} reads ${\cal D} = (1 + \beta \cos \theta)/\sqrt{1-\beta^2}$, and the aberration relation \eqref{aberrnv} simply reduces to
\be\label{EqAberrationCoordinates}
\tilde{\phi} = \phi\,, \qquad \cos\tilde \theta = \frac{\cos \theta + \beta}{1 + \beta \cos \theta}\quad \Rightarrow\quad \sin \tilde \theta = \frac{\sin \theta}{\gamma(1 + \beta \cos \theta)}\,.
\ee
This form of the aberration relation is plotted for various values of $\beta$ in Fig.~\ref{DopAngle}.

It is clear that for directions pointing initially (before the boost) in the hemisphere defined by the boost velocity (that is with $\theta \le \pi/2$), the Doppler shift factor~\eqref{alphadef} satisfies ${\cal D} \ge 1$ for all values of $\beta$ (meaning that the energy is always blueshifted) and tends to infinity when $\beta \to 1$. However the behaviour is different for directions pointing away from the velocity hemisphere (for $\theta > \pi/2$ hence $\cos \theta <0$), since for small $\beta$ it behaves as ${\cal D} = 1 + \beta \cos \theta + {\cal O}(\beta^2)$. In that case for small velocities the Doppler shift factor decreases (the energy is redshifted) as $\beta$ increases, until reaching a minimum ${\cal D}_{\rm min}=\sqrt{1- \cos^2 \theta}$ when $\beta_{\rm min} = -\cos \theta$. For larger values ($\beta\ge \beta_{\rm min}$) the Doppler shift factor increases and the condition ${\cal D}\ge 1$ (blueshifting of energy) is satisfied when $\beta \ge -2 \cos \theta/(1+\cos^2 \theta)$ and the Doppler shift factor eventually diverges toward infinity when $\beta \to 1$. Generically, all directions tend to align with the boost direction when $\beta \to 1$, as one finds in that limit $\tilde \theta \to 0$ and ${\cal D}^{-1} \to 0$.

The previous limits for ${\cal D}^{-1}$ and $\tilde\theta$ are not valid for the special point $\theta = \pi$ (when the initial direction is exactly the opposite of the boost velocity) for which $\tilde\theta=\pi$ for all $\beta$ and ${\cal D}^{-1} \to \infty$ when $\beta \to 1$, hence the convergence $\tilde \theta \to 0$ is not uniform as is manifest in the bottom right panel of Fig.~\ref{DopAngle}, and nor is the convergence ${\cal D}^{-1} \to 0$. In fact, it is possible to check that the asymptotic behaviour when $\beta \to 1$ for the Doppler shift factor is ${\cal D}^{-2} \to 2\delta_D(\cos \theta + 1)$ where $\delta_D$ is the Dirac distribution. This is required because the total solid angle must be conserved and from~\eqref{TransdOmega} we must always have $\int {\cal D}^{-2} \dd \Omega = 4\pi$. With the adapted coordinates we find indeed that $2 \pi\int_{-1}^{-1+\epsilon} (1-\beta^2)/(1+\beta \cos \theta)^2\dd \cos \theta = 2\pi(1+\beta) \epsilon/(1-\beta + \beta \epsilon) \to 4 \pi $  in the limit $\beta\rightarrow 1$ which proves the presence in the limit of a Dirac distribution in the direction opposite to the boost velocity ($\theta = \pi$).


\subsection{Transformation of polarization}\label{boost0}

We now detail how a given polarization tensor in a reference frame is perceived (i.e.\ measured) in a boosted frame. 
The transformation rule of the GW polarization tensor is similar to the transformation rule of the polarization vector for light which is detailed in appendix \ref{Cyril}. 

The gauge condition \eqref{Gaugeepsilon} does not fully determine the polarization, and all polarization tensors of the form $\epsilon_{\mu\nu} + 2\xi_{(\mu} k_{\nu)}$, with $k^\mu \xi_\mu =0$ are equally valid.
We find that to ensure the extra Coulomb gauge condition $\tilde{u}^\mu \tilde{\epsilon}_{\mu\nu}=0$, we need to choose
\be
\tilde{\epsilon}_{\mu\nu} = \epsilon_{\mu\nu} + \xi_\mu \hat{k}_\nu + \xi_\nu \hat{k}_\mu\,,
\ee
with $\hat{k}_\mu \equiv k_\mu / \tilde E$ and
\be
\xi_\mu = \frac{1}{2} \left(\epsilon_{\sigma\nu} \tilde u ^\sigma \tilde u ^\nu \right)\hat{k}_\mu + \epsilon_{\mu\nu} \tilde u ^\nu = \left(\frac{1}{2}\gamma^2 v^\sigma v^\nu\epsilon_{\sigma\nu} \right)\hat{k}_\mu+ \gamma v^\nu \epsilon_{\nu \mu}\,.
\ee
Explicitly, one has\footnote{This transformation rule can be written using the screen projection operator~\eqref{Eq:ScreenUse} as $\tilde{\epsilon}_{\mu\nu} = \tilde{S}_{\mu}^{\,\,\alpha}\tilde{S}_{\nu}^{\,\,\beta} \epsilon_{\alpha\beta}$, similarly to the vector case. The structure of the projection is in all aspects similar to the transformation under a boost of the tensor valued distribution function describing statistically unpolarized photon background~\cite{Challinor:2000as,Tsagas:2007yx,Pitrou:2019hqg}.}
\be
\tilde{\epsilon}_{\mu\nu}  = \epsilon_{\mu\nu} + 2\tilde{u}^\alpha \epsilon_{\alpha(\mu} (\tilde{u}_{\nu)} - \tilde{n}_{\nu)}) + \epsilon_{\alpha\beta}\tilde{u}^\alpha\tilde{u}^\beta(\tilde{u}_{\mu} - \tilde{n}_{\mu})(\tilde{u}_{\nu} - \tilde{n}_{\nu})\,.
\ee
The components\footnote{The tilde stretches over the tensor name and its components to emphasize that not only it is not the same tensor so as to satisfy the Coulomb gauge condition, but its components are also not read in the same basis.} $\widetilde{\epsilon_{ij}} \equiv \tilde{\epsilon}_{\mu\nu}\tilde{e}_i^\mu \tilde{e}_j^\nu$ are found like for the vector case in appendix \ref{Cyril}. One eventually finds 
\be\label{Eq:guessepsilon}
\widetilde{\epsilon_{ij}} = R_{i}^{\,\,k} R_{j}^{\,\,l} \epsilon_{kl}\,,
\ee
or explicitly
\be
\label{eq:pol_trans}
\widetilde{\epsilon_{ij}} = \left[\delta_{i}^k-\frac{1}{\mathcal{D}}\left((\gamma-1)\hat{v}_i + \gamma \beta n_i\right)\hat{v}^k\right] \left[\delta_{j}^l-\frac{1}{\mathcal{D}}\left((\gamma-1)\hat{v}_j + \gamma \beta n_j\right)\hat{v}^l\right]\epsilon_{kl}\,.
\ee
This is consistent with the results of~\cite{Bonvin:2022mkw}, which showed that, at linear order in velocities, the effect of a boost is to rotate the polarization tensor. Here we show that this result is valid at any order in velocity. The direction of propagation also undergoes the same rotation [see eq.~\eqref{ntildeRni}], hence when projecting the rotated polarization on a screen orthogonal to the aberrated direction, the only effect one can have is a rotation of polarization modes, as we now detail. 

First, the basis ${\bm e}_{\theta},{\bm e}_{\phi}$ used to define helicities at the aberrated direction $\widetilde{n_i}$ is not equal to the one at the initial direction $n_i$ rotated by this same rotation $R_{ij}$: 
\be\label{EqInequality}
\tilde{e}^i_{\theta}(\tilde{{\bm n}}) \neq R^i_{\,\,j} {e}^j_{\theta}({{\bm n}})\,,\quad \quad \tilde{e}^i_{\phi}(\tilde{{\bm n}}) \neq R^i_{\,\,j} {e}^j_{\phi}({{\bm n}})\,.
\ee
Instead, defining $R_{\hat d}(\varphi)$ as the rotation around an axis $\hat{d}$ by an angle $\varphi$, there exists a spin phase between ${e}^i_{\theta,\phi}(\tilde{{\bm n}})$ and $R^i_{\,\,j} {e}^j_{\theta,\phi}({{\bm n}})$ (see e.g.\ C10 of \cite{Boyle:2015nqa} or appendix A of \cite{Challinor:2005jy})
\be
\tilde{e}^i_{\theta}(\tilde{{\bm n}})  = [R_{\widetilde{{\bm n}}}(-\delta)]^i_{\,\,j} R^j_{\,\,k} {e}^k_{\theta}({{\bm n}}) = R^i_{\,\,j} [R_{{\bm n}}(-\delta)]^j_{\,\,k}  {e}^k_{\theta}({{\bm n}})\,,
\ee
and similarly for $e_\phi$. 
If we introduce an helicity basis 
\be
{\bm e}_\pm = \frac{1}{\sqrt{2}}({\bm e}_\theta \mp {\rm i} {\bm e}_{\phi})\,,
\ee
then it transforms as
\be\label{TransfoHelicity}
\tilde{e}^i_{\pm}(\tilde{{\bm n}})  = {\rm e}^{\mp {\rm i}\delta} R^i_{\,\,k} {e}^k_{\pm}({\bm n})\,.
\ee
Therefore, from \eqref{Eq:guessepsilon} and the previous transformation of the helicity basis, the polarization components in the helicity basis 
\be
\epsilon_\pm ({\bm n})\equiv \epsilon_{ij} {e}^i_{\mp}({\bm n}){e}^j_{\mp}({\bm n})
\ee
transform according to
\be\label{TransfoHelicityComponents1}
\tilde \epsilon_{\pm}(\tilde{{\bm n}}) \equiv \widetilde{\epsilon_{ij}}  \tilde{e}^i_{\mp}(\tilde{{\bm n}}) \tilde{e}^j_{\mp}(\tilde{{\bm n}}) = {\rm e}^{\pm 2{\rm i}\delta} \epsilon_{\pm}({\bm n})\,.
\ee

It is often more practical to use another basis to extract the polarization components, namely to consider the components
\begin{subequations}
\begin{align}\label{Defpluscross}
\epsilon_{(+)}({\bm n}) &\equiv \frac{1}{2}\left[\epsilon_+({\bm n}) + \epsilon_-({\bm n})\right]= \epsilon_{ij} {e}^i_\theta({\bm n}) {e}^j_\theta ({\bm n}) = - \epsilon_{ij} {e}^i_\phi({\bm n}) {e}^j_\phi ({\bm n})\,,\\
\epsilon_{(\times)}({\bm n}) &\equiv  \frac{1}{2 \ii}\left[\epsilon_+({\bm n}) - \epsilon_-({\bm n})\right]= \epsilon_{ij} {e}^i_\theta({\bm n}) {e}^j_\phi ({\bm n}) \,.
\end{align}
\end{subequations}
which transform as 
\begin{subequations}
\begin{align}\label{spinphasecossineffect}
\tilde \epsilon_{(+)}(\tilde{\bm n}) &= \cos(2 \delta) \epsilon_{(+)}({\bm n}) - \sin(2 \delta) \epsilon_{(\times)}({\bm n})\,,\\
\tilde \epsilon_{(\times)}(\tilde{\bm n}) &= \sin(2 \delta)\epsilon_{(+)}({\bm n}) + \cos(2 \delta) \epsilon_{(\times)}({\bm n})\,.
\end{align}
\end{subequations}

Finally, note that in the particular case where the azimuthal direction of the spherical coordinates system is aligned with the boost velocity direction as in section~\ref{SecAdapted}, the effect of the rotation $R_i^{\,\,j}$ on the direction $n_j$ consists in a transport along the meridian circles ($\phi = {\rm cst}$), and the effect on the polarization $\epsilon_{ij}$ consists in a parallel transport along these circles. Since by construction ${\bm e}_\theta$ and ${\bm e}_\phi$ are also parallel transported along the meridian circles, the spin phase $\delta$ vanishes and eqs.~\eqref{EqInequality} become equalities.

\section{Effect of a boost on a gravitational wave}\label{boostwave}

We now have all the ingredients to study the effect of a boost on a GW. We recall that the results of the previous section are general: the tilde and not tilde frames are related by a boost with velocity ${\bm v}$. 
We now consider a set of sources and observers defined in the following way:
\begin{itemize}
    \item at a given point of spacetime we have two sources. The first source is a reference source with frame $\mathcal{S}$ (whose 4-velocity is $u_s^\mu$) and the second moves with velocity $\bv_s$ with respect to the reference source (at the moment of emission), hence defining the associated boosted frame $\widetilde{\mathcal{S}}$ with 4-velocity  $\tilde u_s^\mu$. 
    The directions of a given eikonal momentum $k^\mu$ defined in~\eqref{kdecomposition} are $n^s_i$ for the $\mathcal{S}$ frame and $\widetilde{n^s_i}$ for the $\widetilde{\mathcal{S}}$ frame, and they are related by the aberration rotation $\widetilde{n^s_i} = R^{s\,j}_{i} n^s_j$ where $R^s_{ij} \equiv R_{ij}(v^s_k,n^s_k)$. 
    
    \item at a given point of spacetime we have two observers looking at the same source. The first observer is a reference observer, whose frame is $\mathcal{O}$ (whose 4-velocity is $u_o^\mu$), and the second observer is boosted with velocity $\bv_o$ with respect to the reference observer (at the moment of reception), hence defining a frame $\widetilde{\mathcal{O}}$ with 4-velocity  $\tilde u_o^\mu$. 
    The directions of a given eikonal momentum $k^\mu$ defined in~\eqref{kdecomposition} are $n^o_i$ for the $\mathcal{O}$ frame and $\widetilde{n^o_i}$ for the $\widetilde{\mathcal{O}}$ frame, and they are related by the aberration rotation $\widetilde{n^o_i} = R^{o\,j}_{i} n^o_j$ where $R^o_{ij} \equiv R_{ij}(v^o_k,n^o_k)$. 
    
\end{itemize}

We want to compare to the reference case $(\mathcal{S},\mathcal{O})$ the two following situations:
\begin{itemize}
\item situation A: a signal is emitted in $\widetilde{\mathcal{S}}$ and received by $\mathcal{O}$\,;
\item situation B: a signal is emitted in $\mathcal{S}$ and received by $\widetilde{\mathcal{O}}$\,.
\end{itemize}
In particular we want to relate the strain measured by the observer in situation A to the reference case, and similarly for situation B. 
For simplicity we first assume throughout this section that the amplitude emitted by the source is isotropic in its frame, that is $\Bsource$ (defined in~\eqref{DefB})  is an intrinsic constant of the source and does not depend on the direction of emission. The case of realistic sources with an anisotropic emission is then detailed in section~\ref{SecObsImplications}.

\subsection{Effect on redshift}\label{SecRedshift}

A signal emitted by the reference source, and measured by the reference observer is seen with a reference redshift defined by
\begin{equation}
\label{eq:zref}
1+z_{\rm ref} = \frac{k_{\mu}u_s^{\mu}}{k_{\mu}u_o^{\mu}}= \frac{E_s}{E_o}\,.
\end{equation}

From the transformation of energy~\eqref{TRuleE}, let us define Doppler shift factors  between the boosted observer and the reference observer, and between the boosted source and the reference source respectively as
\be\label{DefDoDs}
\Do \equiv \frac{k_\mu \tilde u^\mu_o}{k_\mu u^\mu_o} =\frac{\widetilde{E}_o}{E_o} \,,\qquad \Ds \equiv \frac{k_\mu \tilde u^\mu_s}{k_\mu u^\mu_s}=\frac{\widetilde{E}_s}{E_s}\,.
\ee
In situation A and B, the redshift transforms respectively as
\begin{align}
\label{ztildeDsDo}
\mbox{A:}\quad 1+\tilde z \equiv \frac{k_\mu \tilde u^\mu_s}{k_\mu u^\mu_o} = \Ds (1+z_{\rm ref})\,,\qquad
\mbox{B:}\quad 1+\tilde z \equiv \frac{k_\mu u^\mu_s}{k_\mu \tilde u^\mu_o} = {\cal D}^{-1}_o (1+z_{\rm ref})\,.
\end{align}
In the most general situation in which both source and observer move with respect to the reference source and reference observer, respectively, we have
\be\label{JointDoppler}
1+\tilde z \equiv \frac{k_{\mu} \tilde u_s^{\mu}}{k_{\mu} \tilde u_o^{\mu}}=\frac{\widetilde{E}_s}{\widetilde{E}_o} ={\cal D}^{-1} \frac{E_s}{E_o} ={\cal D}^{-1} (1+z_{\rm ref})\,,\qquad \mbox{with}\quad {\cal D} \equiv \frac{\Do}{\Ds}\,.
\ee
The joint Doppler factor ${\cal D}$ is a central quantity in the transformation rules. When the transformation rule of a given  quantity depends only on $\cal D$, the transformation is said to be symmetric in the roles of the source and observer, since at linear order in velocities we have ${\cal D} \simeq 1+ (\bm{v}_o - \bm{v}_s)\cdot \bm{n}$.

If the observer moves radially, along the direction $n^i$, we can define an observer radial velocity $v_o^i = v_o^{\rm rad} n^i$. Similarly if the source moves radially in the direction of $n^i$ we define $v_s^i = v_s^{\rm rad} n^i$. 
The Doppler factors then take the simpler form
\begin{equation}\label{SimpleD}
\Do = \sqrt{\frac{1+v_o^{\rm rad}}{1-v_o^{\rm rad}}}\,,\qquad \Ds = \sqrt{\frac{1+v_s^{\rm rad}}{1-v_s^{\rm rad}}}\,.
\end{equation}
If the observer moves toward the source, $v_o^{\rm rad}>0$, and for relativistic velocities we can use the approximation $\Do\sim 2\gamma_o$. If the source moves toward the observer, $v_s^{\rm rad}<0$, and for relativistic velocities we have $\Ds^{-1}\sim 2\gamma_s$.

\subsection{Effect on distances}\label{secdistances}

As seen in section~\ref{sec:amplitude_gen}, distances depend on the ratio of surface elements $\dd A$ and solid angles $\dd\Omega$.
The area of the cross-section of the beam at the observer is well-defined independently of the four-velocity of the observer. 
Indeed, if one observer with $u^\mu$ defines two separation vectors $\xi_{1,2}^\mu$ such that $\xi_{1,2}^\mu u_\mu = k_\mu \xi_{1,2}^\mu = 0$ (i.e.\ $\xi_{1,2}^\mu$ are spatial for this observer and orthogonal to the wave propagation), then another observer with $\tilde u^\mu$ would define two related separation vectors by $\tilde{\xi}_{1,2}^\mu =  \xi_{1,2}^\mu + \alpha_{1,2} k^\mu$, with $\alpha_{1,2}$ chosen such that $\tilde{\xi}_{1,2}^\mu \tilde{u}_\mu = \tilde{\xi}_{1,2}^\mu k_\mu =0$. It is clear that $\xi_{1,2}^\mu \xi_{1,2}^\nu g_{\mu\nu} = \tilde \xi_{1,2}^\mu \tilde \xi_{1,2}^\nu g_{\mu\nu}$, hence shapes and areas are preserved. Therefore the concept of beam shape and area, wherever the beam is not singular, is observer independent (see also section 3.4 of \cite{1992grle.bookS} or section 2.1.2 of \cite{Fleury:2015hgz}).

On the other hand, in order to measure the size of the beam where it converges in terms of a solid angle $\dd\Omega$  it is necessary to take into account the 4-velocity of the frame in which angles are measured.  Let us assume that, at that point, we have two observers with velocities $u^{\mu}$ and $\tilde u^{\mu}$. The solid angles in the two cases are related via
\be\label{TransdOmega}
\frac{\dd\Omega}{\dd\widetilde{\Omega}}={\cal D}^2=\left(\frac{k_{\mu}\tilde u^{\mu}}{k_{\mu}u^{\mu}}\right)^2\,.
\ee
Indeed, using natural spherical coordinates associated with the zenith direction ${\bm{ \hat v}}$ as in section~\ref{SecAdapted}, $\dd \Omega = \sin \theta \dd \theta \dd \phi$. From \eqref{EqAberrationCoordinates}, we find $\sin \tilde\theta \dd \tilde \theta / \dd \theta = {\cal D}^{-2}\sin \theta$, with ${\cal D} = \gamma(1 + \beta \cos \theta)$, and therefore $\sin \tilde {\theta} \dd \tilde\theta  \dd \tilde \phi = {\cal D}^{-2}\sin \theta \dd \theta \dd \phi$.

This change of solid angles holds when considering two different sources with Doppler shift $\Ds$, emitting the same beam toward the observer, and the solid angle at emission is either $\dd \Omega_s$ or $\dd \widetilde{\Omega}_s$ depending on the source frame, or when considering two different observers with Doppler shift $\Do$ sending a fictitious beam towards the same source, and the solid angle of the fictitious beam they send is either $\dd \Omega_o$ or $\dd \widetilde{\Omega}_o$ depending on the observer frame,\footnote{Note that while in the case of GW sources the notion of solid angle at the observer is a pure mathematical construction, in other situations such solid angles are physically relevant, e.g.\ if one observes galaxies, whose area are seen under solid angle $\dd\Omega_o$.} hence
\be\label{TransdOmega2}
\frac{\dd\widetilde{\Omega}_o}{\dd\Omega_o }=\Do^{-2}\,,\qquad \frac{\dd\widetilde{\Omega}_s}{\dd\Omega_s }=\Ds^{-2}\,.
\ee
With this we can determine how distances are affected in situation A and B.

\subsubsection{Moving source (situation A) :  $\Do=1, \Ds\neq 1$}

We want to relate the distance $\chi(\mathcal{S} \rightarrow \mathcal{O})$, from the source frame $\mathcal{S}$ to the observer $\mathcal{O}$, to the distance $\chi(\widetilde{\mathcal{S}} \rightarrow \mathcal{O})$, from the source frame $\widetilde{\mathcal{S}}$ to the observer $\mathcal{O}$. Using~\eqref{TransdOmega2} we obtain
\begin{equation}\label{eq:chis}
\chi(\widetilde{\mathcal{S}}\rightarrow \mathcal{O})=\sqrt{\frac{\dd A_o}{\dd\widetilde{\Omega}_s}}=\sqrt{\frac{\dd A_o}{\dd{\Omega}_s}}\sqrt{\frac{\dd{\Omega}_s}{\dd\widetilde{\Omega}_s}}=\Ds {\chi}({\mathcal{S}}\rightarrow \mathcal{O})\,.
\end{equation}
The distance $\chi$ is therefore directly affected by the source velocity.

On the other hand, the angular distance $D_A$ at which both sources are seen by the reference observer is invariant, in accordance to \eqref{Eq.chitoDA}, that is (notice that here the arrow goes to the left because in the definition of the angular diameter distance we are considering a virtual beam going from the observer towards the source)
\be\label{TDAA}
D_A(\widetilde{\mathcal{S}}\leftarrow \mathcal{O}) = D_A(\mathcal{S}\leftarrow \mathcal{O}) \,.
\ee
The  transformation of $D_L$ is found from the relations~\eqref{DefDL} and we get
\begin{equation}\label{TDLA}
\frac{D_L(\widetilde{\mathcal{S}}\rightarrow \mathcal{O})}{D_L({\mathcal{S}}\rightarrow \mathcal{O})} = \left(\frac{1+\tilde z}{1+z_{\rm ref}}\right) \frac{\chi(\widetilde{\mathcal{S}}\rightarrow \mathcal{O})}{\chi({\mathcal{S}}\rightarrow \mathcal{O})} = \Ds^{2}\,.
\end{equation}

\subsubsection{Moving observer (situation B):  $\Ds=1, \Do\neq 1$}

We want to relate the distance ${\chi}$ measured from the source $\mathcal{S}$ to $\widetilde{\mathcal{O}}$, to the reference distance $\chi$, separating the reference source and observer,  $\mathcal{S}$ to $\mathcal{O}$. Since $\chi$ does not depend on the observer velocity by definition 
\be\label{chiunchanged}
{\chi}(\mathcal{S}\rightarrow \widetilde{\mathcal{O}}) = \chi(\mathcal{S}\rightarrow \mathcal{O})\,.
\ee

However, in this situation B, the angular distance at which both observers 
see the reference source is not the same. We recall that the solid angles at which they see a source is modified by aberration, according to \eqref{TransdOmega2} evaluated at the observer. Hence, using~\eqref{TransdOmega2}, the respective angular distances are related by 
\begin{equation}\label{TDAB}
D_A({\mathcal{S}}\leftarrow \widetilde{\mathcal{O}})=\sqrt{\frac{\dd A_s}{\dd\widetilde{\Omega}_o}}=\sqrt{\frac{\dd A_s}{\dd{\Omega}_o}}\sqrt{\frac{\dd{\Omega}_o}{\dd\widetilde{\Omega}_o}}= \Do {D}_A (\mathcal{S}\leftarrow {\mathcal{O}})\,.
\end{equation}
The transformation of the luminosity distance is obtained from definition~\eqref{DefDL} and we get
\begin{equation}\label{TDLB}
\frac{D_L(\mathcal{S}\to \widetilde{\mathcal{O}})}{D_L(\mathcal{S}\to {\mathcal{O}})} = \left(\frac{1+\tilde z}{1+z_{\rm ref}}\right) \frac{\chi(\mathcal{S}\to \widetilde{\mathcal{O}})}{\chi(\mathcal{S}\to {\mathcal{O}})} = \Do^{-1}\,.
\end{equation}

\subsubsection{Summary}

We can now consider a situation where both the source and the observer move with respect to the reference source and observer respectively. We need only to combine the transformation relations of the previous sections and we find
\begin{subequations}
\begin{empheq}[box=\widefbox]{align}
\chi(\widetilde{\mathcal{S}}\rightarrow \widetilde{\mathcal{O}})&=\Ds\, \chi(\mathcal{S}\rightarrow {\mathcal{O}})\,,\label{MegaTchi}\\
D_A(\widetilde{\mathcal{S}}\leftarrow \widetilde{\mathcal{O}})&=\Do\, {D_A}(\mathcal{S}\leftarrow {\mathcal{O}})\,,\label{MegaTDA}\\
D_L(\widetilde{\mathcal{S}}\rightarrow \widetilde{\mathcal{O}})&=\frac{\Ds^2}{\Do} {D_L}(\mathcal{S}\rightarrow {\mathcal{O}})\label{MegaTDL}\,.
\end{empheq}
\end{subequations}

We see that all three distances transform in an asymmetric way under a boost of the source and of the observer, i.e.\ the Doppler factors $\Ds^{-1}$ and $\Do$ appear with different powers such that the transformations rules cannot be expressed in terms of the joint Doppler factor ${\cal D}$. This asymmetry, which had already been found at linear order in the velocity~\cite{Bonvin:2005ps,Hui:2005nm}, does not break any equivalence principle, but it is simply linked to the fact that situation A and B are not physically equivalent, as we will discuss in more detail in section~\ref{sec:interpretation}. We also check that the ratio of $\chi/D_A$ transforms with a factor ${\cal D}^{-1}$, in the same way as $(1+z)$ in~\eqref{JointDoppler}. From this, it follows that $\chi/D_A \propto (1+z)$ which is consistent with the relation~\eqref{Eq.chitoDA}.\footnote{This can even be viewed as a proof of~\eqref{Eq.chitoDA} since both distances must agree for small redshifts and separations, hence the constant of proportionality must be unity and therefore $\chi/D_A = (1+z)$.}

\subsection{Effect on amplitude and energy density}\label{secamplitude}

We now deduce how the strain is transformed for a source emitting isotropically, from~\eqref{ampl}, using that in such a case $\Bsource$ is an intrinsic property of the source. 
From the transformation rule~\eqref{MegaTchi} we find that when the source moves with respect to the reference source, the strain is modified as
\be\label{strainrule1}
{H}(\widetilde{\mathcal{S}}\rightarrow {\mathcal{O}})=\Ds^{-1} H(\mathcal{S}\rightarrow \mathcal{O})\,.
\ee
For a source moving radially towards the observer $\Ds<1$ and the strain is therefore amplified by the source velocity. In the case of relativistic velocities, this amplification becomes $2\gamma_s$.
On the other hand, when the observer moves with respect to the reference observer the strain does not change
\be\label{strainrule1b}
{H}({\mathcal{S}}\rightarrow \widetilde{\mathcal{O}})=H(\mathcal{S}\rightarrow \mathcal{O})\,.
\ee
This is expected since in the latter case both observers measure the same scalar field (the GW amplitude) at the same point of spacetime, hence they must agree on the observed amplitude, whereas in the former case an observer is comparing two physically different signals emitted by different sources. 

We are now in position to deduce how the total energy density of the signal received by the observer is transformed. The energy which appears in the expression~\eqref{rho} of the GW energy density is the observed energy, that is $k_\mu u^\mu_o$ for the reference observer and $k_\mu \tilde u^\mu_o$ for the boosted observer. We must also consider that the source emits a GW with a given energy in its frame. From~\eqref{JointDoppler} the energies are related by
\be
\frac{\widetilde{E}_o/\widetilde{E_s}}{E_o/E_s} = \frac{1+ z_{\rm ref}}{1+\tilde{z}} = {\cal D}\,.
\ee
Comparing two sources that are emitting the same energy (as measured in their own proper frame), i.e.\ with $\widetilde{E}_s=E_s$ we obtain
\begin{align}
\label{TEoD}
 \widetilde{E}_o={\cal D}E_o\, .
\end{align}
Hence the energy measured transforms symetrically.

Using \eqref{strainrule1}, \eqref{strainrule1b} for the transformation of the strain, and~\eqref{TEoD} for the transformation of the observed energy, we obtain the following relations for the energy  density and the strain, in the general case where both the source and the observer move with respect to the reference situation
\begin{subequations}
\begin{empheq}[box=\widefbox]{align}
{H}(\widetilde{\mathcal{S}}\rightarrow {\widetilde{\mathcal{O}}})&=\Ds^{-1} H(\mathcal{S}\rightarrow \mathcal{O})\,,\\
{\rho}_{{\rm GW}}(\widetilde{\cal S} \rightarrow \widetilde{\cal O}) &= \Do^2\, \Ds^{-4} \rho_{{\rm GW}}({\cal S} \rightarrow {\cal O})\,.\label{rhoT1}
\end{empheq}
\end{subequations}
We recover that the energy density transforms as $D_L^{-2}$ does, in agreement with the definition of the luminosity distance. The asymmetry in the source and observer velocity, i.e.\ the fact that the Doppler factors $\Do$ and $\Ds^{-1}$ appear with different powers in the transformation rule, is directly due to the fact that the strain scales as the inverse of the distance $\chi$ and that distances do not transform in a symmetric way under a boost of source and observer.

\subsection{Interpretation of the asymmetry in the distance and amplitude}
\label{sec:interpretation}

Let us recall the definitions of distances $\chi$ and $D_A$. The distance $\chi$ is the distance defined by the source when the signal is received by the observer.  On the other hand, the angular diameter distance is the distance $\chi$ defined by 
an observer sending a fictitious signal toward the past, and which would be received by the source at emission time. It corresponds to a distance defined by the observer when the fictitious signal reaches the source (hence at the emission time).

\subsubsection{Galilean relativity}
Let us focus first on effects linear in $v/c$, and consider a flat background. This corresponds to restricting to Galilean relativity only. Let us also assume that in the reference situation the source and the observer have the same velocity, hence $z_{\rm ref}=0$. In that reference situation, the observer is at a distance $d_{\rm ref}$ both when the signal is emitted and when the signal is received. However in situation A, if the moving source moves toward the observer (compared to the reference situation), the distance at emission is still $d_{\rm ref}$ but it is reduced to $d_{\rm ref}(1-v/c)$ when the signal is received. The former distance corresponds to the angular distance $D_A(\widetilde{\mathcal{S}} \leftarrow \mathcal{O})$, whereas the latter distance is the distance of propagation of the signal $\chi(\widetilde{\cal S} \rightarrow {\cal O})$. Since, for the moving source, the distance at reception is reduced, the dilution of the wave amplitude from propagation is also reduced and the signal is enhanced. 

In situation B, the observer moves towards the source (compared to the reference situation). The distance at emission must be $d_{\rm ref}(1+v/c)$ so that it is reduced to $d_{\rm ref}$ at reception due to the motion of the observer toward the source. The former distance is $D_A({\mathcal{S}} \leftarrow \widetilde{\mathcal{O}})$ whereas the latter distance is the propagation distance $\chi(\cal{S} \rightarrow \widetilde{\cal{O}})$. Hence in this case, the distance at reception is the same for the moving observer and the reference observer, meaning that the wave amplitude is not affected by the observer velocity.

From this, we see that even though in both situations the redshift induced by the motion ($z=-v/c$) is the same (be it the source moving toward the reference observer, or the observer moving towards the reference source), the situations are not symmetric in terms of distances. For instance from the discussion above we have
\be
\chi(\widetilde{\mathcal{S}} \rightarrow {\mathcal{O}}) = \chi(\mathcal{S} \rightarrow \widetilde{\mathcal{O}}) - d_{\rm ref} \frac{v}{c}\,.
\ee
In other words, in situation B the source and the observer are separated by a propagation distance $\chi(\mathcal{S} \rightarrow \widetilde{\mathcal{O}})=d_{\rm ref}$ when the signal is received, whereas in situation A they are separated by a propagation distance $\chi(\widetilde{\mathcal{S}} \rightarrow {\mathcal{O}})=d_{\rm ref}(1-v/c)$ when the signal is received. Therefore the amplitude of the GW is not the same in these two physically different situations. Note that this also explains why the luminosity distance $D_L$ is not affected in the same way by the source and observer velocities, as shown in~\cite{Bonvin:2005ps,Hui:2005nm}.\footnote{In~\cite{Kaiser:2014jca} it was argued that to solve the asymmetry between the impact of source and observer velocity on distances it is necessary to account for the evolution of velocities with time and to include the effect of gravitational redshift. Here we show however that this asymmetry is real and does not violate the equivalence principle.} If we want situation B to be exactly as situation A up to a Galilean transformation, we must also shift the observer worldline toward the source by a displacement $d_{\rm ref}\,v/c$, and this would enhance the signal observed as in situation A with the rule~\eqref{strainrule1}.

\subsubsection{Special relativity}
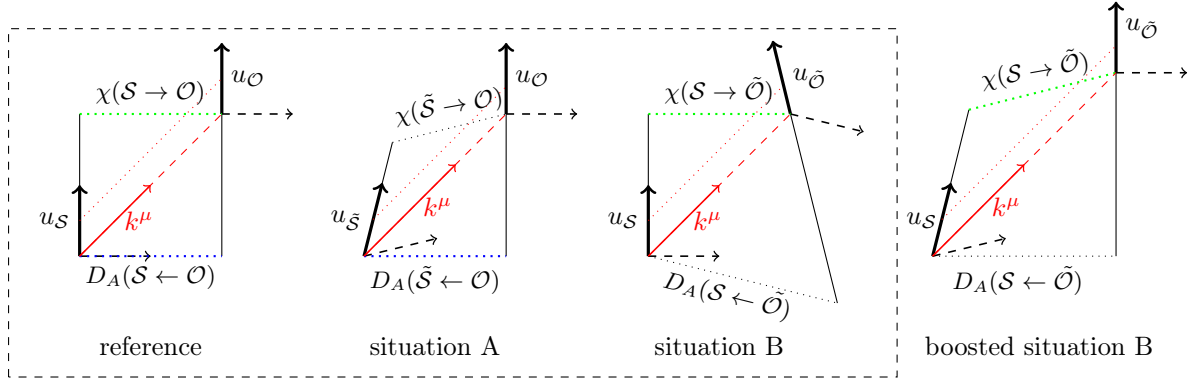
\begin{figure}[!htb]
\begin{tikzpicture}[scale=0.94]
\draw[->, line width=1.2pt] (0,0) -- (0,1) ;
\draw[] (0,0) -- (0,2) ;
\node[left] at (0,0.5) {$u_{\cal{S}}$};
\draw[->, dashed, line width=0.6pt] (0,0) -- (1,0) ;

\draw[->, line width=1.2pt] (2,2) -- (2,3) ;
\draw[] (2,0) -- (2,2) ;
\node[right] at (2,2.5) {$u_{\cal{O}}$};
\draw[->, dashed, line width=0.6pt] (2,2) -- (3,2) ;

\draw[dotted, line width=0.8pt, color=blue] (0,0) -- (2,0) ;
\draw[dotted, line width=0.8pt, color=green] (0,2) -- (2,2) ;
\draw[dashed, color=red] (0,0) -- (2,2) ;
\draw[->, color=red, line width=0.6pt] (0,0) -- (1,1) ;
\node[right] at (0.5,0.5) {${\color{red} k^\mu}$};
\draw[dotted, color=red] (0,0.5) -- (2.,2.5) ;
\node[below] at (1,0) {\footnotesize $D_A(\cal{S} \leftarrow \cal{O})$};
\node[above] at (1,2) {\footnotesize $\chi(\cal{S} \rightarrow \cal{O})$};
\node[below] at (1,-1) {reference};

\draw[->, line width=1.2pt] (4,0) -- (4.258,1.03) ;
\draw[] (4,0) -- (4.4,1.6) ;
\node[left] at (4.1,0.5) {$u_{\tilde{\cal{S}}}$};
\draw[->, dashed, line width=0.6pt] (4,0) -- (5.03,0.25) ;

\draw[->, line width=1.2pt] (6,2) -- (6,3) ;
\draw[] (6,0) -- (6,2) ;
\node[right] at (6,2.5) {$u_{\cal{O}}$};
\draw[->, dashed, line width=0.6pt] (6,2) -- (7,2) ;

\draw[dotted, line width=0.8pt, color=blue] (4,0) -- (6.,0) ;
\draw[dotted] (4.4,1.6) -- (6,2) ;
\draw[dashed, color=red] (4,0) -- (6.,2.) ;
\draw[->, color=red, line width=0.6pt] (4,0) -- (5.29,1.29) ;
\node[right] at (4.7,0.7) {${\color{red} k^\mu}$};
\draw[dotted, color=red] (4.13,0.516) -- (6.,2.387) ;
\node[below] at (5,0.) {\footnotesize$D_A(\tilde{\cal{S}} \leftarrow \cal{O})$};
\node[above,rotate=10] at (5.25,1.8) {\footnotesize $\chi(\tilde{\cal{S}} \rightarrow \cal{O})$};
\node[below] at (5,-1) {situation A};

\draw[->, line width=1.2pt] (8,0) -- (8,1) ;
\draw[] (8,0) -- (8,2) ;
\node[left] at (8,0.5) {$u_{\cal{S}}$};
\draw[->, dashed, line width=0.6pt] (8,0) -- (9,0) ;

\draw[->, line width=1.2pt] (10.,2) -- (9.75,3.03) ;
\draw[] (10.66,-0.66) -- (10,2) ;
\node[right] at (9.9,2.5) {$u_{\tilde{\cal{O}}}$};
\draw[->, dashed, line width=0.6pt] (10,2) -- (11.03,1.75) ;

\draw[dotted] (8,0) -- (10.66,-0.66) ;
\draw[dotted,  line width=0.8pt, color=green] (8,2) -- (10,2) ;
\draw[dashed, color=red] (8,0) -- (10,2) ;
\draw[->, color=red, line width=0.6pt] (8,0) -- (9,1.) ;
\node[right] at (8.5,0.5) {${\color{red} k^\mu}$};
\draw[dotted, color=red] (8,0.5) -- (9.9,2.4) ;
\node[below,rotate=-14] at (9.2,-0.2) {\footnotesize $D_A(\cal{S} \leftarrow \tilde{\cal{O})}$};
\node[above] at (9,2) {\footnotesize $\chi(\cal{S} \rightarrow \tilde{\cal{O}})$};
\node[below] at (9,-1) {situation B};

\draw[->, line width=1.2pt] (12,0) -- (12.258,1.03) ;
\draw[] (12,0) -- (12.51,2.06) ;
\node[left] at (12.2,0.5) {$u_{\cal{S}}$};
\draw[->, dashed, line width=0.6pt] (12,0) -- (13.03,0.25) ;

\draw[->, line width=1.2pt] (14.58,2.58) -- (14.58,3.58) ;
\draw[] (14.58,0.) -- (14.58,2.58) ;
\node[right] at (14.58,3.2) {$u_{\tilde{\cal{O}}}$};
\draw[->, dashed, line width=0.6pt] (14.58,2.58) -- (15.58,2.58) ;

\draw[dotted] (12,0) -- (14.58,0) ;
\draw[dotted, line width=0.8pt, color=green] (12.51,2.06) -- (14.58,2.58) ;
\draw[dashed, color=red] (12,0) -- (14.58,2.58) ;
\draw[->, color=red, line width=0.6pt] (12,0) -- (13.29,1.29) ;
\node[right] at (12.7,0.7) {${\color{red} k^\mu}$};
\draw[dotted, color=red] (12.13,0.516) -- (14.58,2.96) ;

\node[below,rotate=0] at (13.2,0.) {\footnotesize $D_A(\cal{S} \leftarrow \tilde{\cal{O})}$};
\node[above,rotate=10] at (13.5,2.3) {\footnotesize $\chi(\cal{S} \rightarrow \tilde{\cal{O}})$};
\node[below] at (13.5,-1) {boosted situation B};

\draw[dashed, color=black] (-1,-1.7) -- (-1,3.2) -- (11.5,3.2) -- (11.5,-1.7) -- (-1,-1.7);

\end{tikzpicture}
\caption{\label{fig:Galilean} \small Comparison of reference situation with situations A and B in a Minkowski background, with $v/c=0.25$. For each source and observer, we show the velocity unit vector (thick continuous arrow) and the spatial unit vector (thin dashed arrow) which is used to measure distances in the associated frame. The distance measured in the source frame when the reference null geodesic (in red dashed line), with initial wave vector $k^\mu$ (red arrow), is received corresponds to $\chi$, whereas the distance measured in the observer frame when the signal is emitted corresponds to $D_A$. In situation A, $D_A$ is the same as in the reference situation, that is $D_A({\cal S} \leftarrow {\cal O}) =D_A(\widetilde{{\cal S}} \leftarrow {\cal O})$, whereas in situation B, it is $\chi$ which is the same as in the reference situation, that is $\chi(\cal{S} \rightarrow \cal{O}) = \chi(\cal{S} \rightarrow \widetilde{\cal{O}})$. The last panel corresponds to situation B, where we have applied a global boost on both the source and the observer such that the observer moves with the same velocity as the reference observer. It is therefore equivalent to situation B, and it can be seen that in order to be equivalent to situation A, one must consider a different observer whose worldline is closer to the source. The dotted red line is a null geodesic corresponding to a pulse emitted after a given source proper time. In both situations A and B, the observer time difference at which these pulses are received is smaller due to Doppler effect, see eq.~\eqref{dPhidtau2} below.}
\label{fig:Minkowski}
\end{figure}

The next level of interpretation to understand how distances are modified is in special relativity, that is when considering a Minkowski background but without restricting expressions to linear order in $v/c$. We also assume that in the reference situation the source and the observer share the same velocity, such that $z_{\rm ref} = 0$. This configuration is illustrated in Fig.~\ref{fig:Galilean}, where we compare the observer and source worldlines in the reference situation, in situation A and situation B. In this context, the distance $\chi$ reduces to the distance measured in the source frame between the source and the observer when the  signal is received, while the distance $D_A$ is the distance  measured in the observer frame between the source and the observer when the signal is emitted. In the figure we attach Minkowski diagrams to the source position at emission and the observer position at reception. They consist of the velocity vector and the associated orthonormal spatial unit vector used to measure distances. In situation A, $D_A$ is the same as in the reference situation, that is $D_A({\cal S} \leftarrow {\cal O}) = D_A(\widetilde{\cal{S}} \leftarrow \cal{O})$, whereas in situation B, it is $\chi$ which is the same as in the reference situation, that is $\chi(\cal{S} \rightarrow \cal{O}) = \chi(\cal{S} \rightarrow \widetilde{\cal{O}})$. The last panel on the right corresponds to the situation in which we boost the whole Minkowski spacetime in such a way that the observer has the same velocity as the reference observer (while the source moves). The resulting picture is equivalent to situation B, and it can be seen that in order to be equivalent to situation A, one has to consider a different observer with the same velocity whose worldline is closer to the source (see caption of Fig.~\ref{fig:Galilean} for more explanations). Again, as in the Galilean relativity interpretation, this latter modification reduces distances, hence it amplifies the strain so as to obtain results exactly similar to situation A, that is the strain transformation is given by eq.~\eqref{strainrule1} when considering a different observer with a worldline shifted closer to the source.

\subsection{Effect on specific intensity}

The object that transforms symmetrically in the transformations of situation A and B is the specific intensity for an ensemble of sources (or for an extended source, if it exists), since this does not depend on distances. In the following, we write this quantity for the reference situation in which the source frame is $\mathcal{S}$ and the observer frame is $\mathcal{O}$. 
For an extended source (or for an ensemble of sources), the energy density associated to the surface element $\dd A_s$ can be defined adapting the definition (\ref{rho}) as 
\be
\dd\rho_{\rm GW}=\frac{c^2}{64 
\pi G}\frac{\dd\Sigma}{\dd E_s} \dd E_s E_o^2 H^2 \dd A_s\,,
\ee
where $\dd A_s$ is the infinitesimal surface element and $\dd\Sigma/\dd E_s$ is the superficial density per unit of emitted energy. Then using (\ref{amplitude}) and recalling that $\dd A_s=\dd\Omega_o  D_A^2$, we obtain
\be\label{specin}
\frac{\dd\rho_{\rm GW}}{\dd\Omega_o \dd\log E_o}(\mathcal{S}\rightarrow \mathcal{O})=\frac{c^2}{64 
\pi G}\frac{\dd\Sigma}{\dd E_s}\Bsource^2E_s^3  \frac{1}{(1+z_{\text{ref}})^4}\,,
\ee
where $\Bsource$ is defined in \eqref{DefB}, and where we used that $\dd\log E_s = \dd\log E_o$. The observed energy $E_o$ corresponds to an emitted energy $E_{\rm ref} = E_s$ and $E_s/E_o = 1+z_{\rm ref}$. Eq.~\eqref{specin} is the expression of the energy density of particles (photons or gravitons) in a logarithmic band of energy received in an infinitesimal solid angle. It is directly proportional to the energy flux per surface area per solid angle (in a logarithmic band of energy), which is the specific intensity received from the (equivalent) extended source (represented by the blue surface in Fig.~\ref{FigDistances}).

If we repeat the procedure with a boosted source and a boosted observer we obtain
\be\label{specin2}
\frac{\dd\rho_{\rm GW}}{\dd\widetilde{\Omega}_o \dd\log \widetilde{E}_o}(\widetilde{\mathcal{S}}\rightarrow\widetilde{\mathcal{O}})=\frac{c^2}{64 
\pi G}\frac{\dd\Sigma}{\dd \widetilde{E}_s}\Bsource^2 \widetilde{E}_s^3  \frac{1}{(1+\tilde{z})^4}\,.
\ee
Eqs.~\eqref{specin} and \eqref{specin2} do not depend on distances anymore.
In eq.~\eqref{specin2}, the observed energy $\widetilde{E}_o$ corresponds to an emitted energy $E_{\rm ref} = \widetilde{E}_s$ with $\widetilde{E}_s/\widetilde{E}_o = 1+\tilde{z} = \Ds/\Do (1+ z_{\rm ref})$. For a given $E_{\rm ref}$ the observed energy when both the source and the observer move is related to the observed energy in the reference situation by~\eqref{TEoD}, that is with the joint Doppler factor.

The quantity
\be
\frac{\dd\rho_{\rm GW}}{\dd{\Omega}_s \dd \log E_{\rm ref}}(E_{\rm ref}) \equiv \frac{c^2}{(64 \pi G)}E_{\rm ref}^3 \Bsource^2 \frac{\dd\Sigma}{\dd E_{\rm ref}}
\ee
is the specific intensity of the extended source, i.e.\ an intrinsic property of the source. When both the source and the observer move with respect to the reference source and reference observer respectively, the transformation rule of specific intensity can be expressed with the joint Doppler factor~\eqref{JointDoppler} as\footnote{If we define the specific intensity as the energy density per solid angle, and per energy band instead of per logarithmic energy band, then the transformation rule is
\be\label{Tspecific3}
\frac{\dd\rho_{\rm GW}}{\dd\widetilde{\Omega}_o \dd \widetilde{E}_o}(\widetilde{\mathcal{S}}\rightarrow \widetilde{\mathcal{O}})=\mathcal{D}^3\frac{\dd\rho_{\rm GW}}{\dd\Omega_o \dd E_o}(\mathcal{S}\rightarrow \mathcal{O})\,,
\ee}
\be\label{Tspecific1}
\frac{\dd\rho_{\rm GW}}{\dd\widetilde{\Omega}_o \dd\log \widetilde{E}_o}(\widetilde{\mathcal{S}}\rightarrow \widetilde{\mathcal{O}})=\mathcal{D}^4\frac{\dd\rho_{\rm GW}}{\dd{\Omega}_o \dd\log  E_o}({\mathcal{S}}\rightarrow {\mathcal{O}})\,,
\ee
where it is understood that the left hand side is evaluated at $\widetilde{E}_o$, whereas the right hand side is evaluated at the related $E_o$, and both energies are related by~\eqref{TEoD}.

In addition, it is immediate to deduce that the total specific intensity obtained by integrating over all logarithmic energy bands
\be\label{DefSpecific}
\frac{\dd\rho_{\rm GW}}{\dd{\Omega}_o} = \int \frac{\dd\rho_{\rm GW}}{\dd\Omega_o \dd\log E_o} \dd \log E_o\,,
\ee
transforms with the same Doppler factors
\be\label{Tspecific2}
\frac{\dd\rho_{\rm GW}}{\dd\widetilde{\Omega}_o}(\widetilde{\mathcal{S}}\rightarrow \widetilde{\mathcal{O}})=\mathcal{D}^4\frac{\dd\rho_{\rm GW}}{\dd\Omega_o}(\mathcal{S}\rightarrow \mathcal{O})\,,
\ee
since $\dd \ln \widetilde{E}_o = \dd \ln E_o$.
This is compatible with the transformation rules for the total energy density received from a source, namely eqs.~\eqref{rhoT1}, since $\dd \Omega_o$ is invariant in situation A, but is modified in situation B as $\dd \widetilde{\Omega}_o = {\cal D}_o^{-2} \dd {\Omega}_o$ [see eq.~\eqref{TransdOmega2}]. This is another way to see that the ratio of energy over unit solid angle, i.e.\ the total specific intensity, transforms with a factor ${\cal D}^4$ in both situations. In situation A, the total energy density received is enhanced by a factor ${\cal D}_s^{-4}$, but the source is seen through the same solid angle $\dd \Omega_o$, whereas in situation B, the total energy received is enhanced by a factor ${\cal D}_o^2$, but from differential aberration it is seen through a solid angle reduced by a factor ${\cal D}_o^{2}$, such that it transforms with a factor ${\cal D}_o^{4}$. An alternative derivation of these transformation properties based on the graviton distribution function can be found in \cite{Cusin:2022cbb} (see also appendix of \cite{Cusin:2018avf}).\footnote{We stress that in the context of an astrophysical background, where we are collecting at the observer position contributions of a collection of sources along the line of a sight, the contribution of source and observer velocities inside the integral is symmetric when integrating along the line of sight over the physical thickness, or equivalently over the source proper time, see equation (1) of \cite{Pitrou:2019rjz}. However, when the integration variable used is conformal time, the contribution becomes asymmetric since $\dd\tau/\dd\eta$ depends only locally on the source velocity and not on the observer velocity, see eq.~(67) of \cite{Cusin:2017fwz}. }

Finally, as detailed in appendix~\ref{SecEikonalA}, this derivation is in all aspects similar to the case of electromagnetism. If in that case the total specific intensity~\eqref{DefSpecific} is the one of a black body, it is proportional to $T^4$, where $T$ is the temperature, hence from the transformation law~\eqref{Tspecific2} we recover that the observed temperature from a given source transforms as $\widetilde{T}_o = {\cal D} T_o$.

\subsection{Effect on frequency}\label{SecPhase}

From eqs.~\eqref{dPhidtau} and \eqref{z}, the frequencies associated with the phase variations for the observer and the source are related by
\be\label{dPhidtau2}
\left.\frac{\dd\Phi}{\dd \tau}\right|_{o} = \frac{1}{(1+z)} \left.\frac{\dd\Phi}{\dd \tau}\right|_{s} \,.
\ee

Since the phase  is conserved during propagation, the phase difference between two pulses emitted by the source (e.g. two maxima of the emitted wave) is the same as the phase difference when these pulses are received by the observer. Therefore, the infinitesimal proper time between two pulses are related by
\be\label{dPhidtau3}
\dd \tau_{o} = (1+z) \dd \tau_s.
\ee
The situation is perfectly symmetric in both situations A and B described above. More precisely if we start from a reference situation with redshift $z_{\rm ref}$, then any modified redfshift $\tilde z$ induced by the motion of the source (situation A) or of the observer (situation B) with respect to this situation is taken into account by using $(1 + \tilde z)$ from eq.~\eqref{JointDoppler} instead of $(1+z)$ in \eqref{dPhidtau2} and \eqref{dPhidtau3}. The transformation rule is therefore
\be\label{Tdtdt}
\frac{\dd \tilde \tau_o}{\dd \tilde \tau_s} = {\cal D}^{-1}\frac{\dd \tau_o}{\dd \tau_s}\,.
\ee

\section{Realistic waveforms}\label{SecObsImplications}

Until now we have assumed that the source was emitting isotropically a signal. In this section we release this assumption, and we generalise our results to the case in which the source emits a non-isotropic wave, as for example a binary system of compact objects. The main difference with respect to the previous treatment is that now the wave intrinsic amplitude $\Bsource$ (defined in~\eqref{DefB}) is a function of the source direction, which is aberrated  by boosts.

\subsection{Global tetrad frame}\label{Synge}

In this section, we construct a special reference observer, related to a given emitting source, by parallel transporting the source tetrad along a geodesic. We also explain how the same construction can be used to define, given an observer, a special reference source frame, parallel transporting the tetrad of the observer back to the source. These constructions are useful to relate components of the wave at source and observer position.

\subsubsection{Synge source and reference observer}\label{SecSyngeSource}

\begin{figure}[!htb]
\centering
\includegraphics[width=0.6\textwidth]{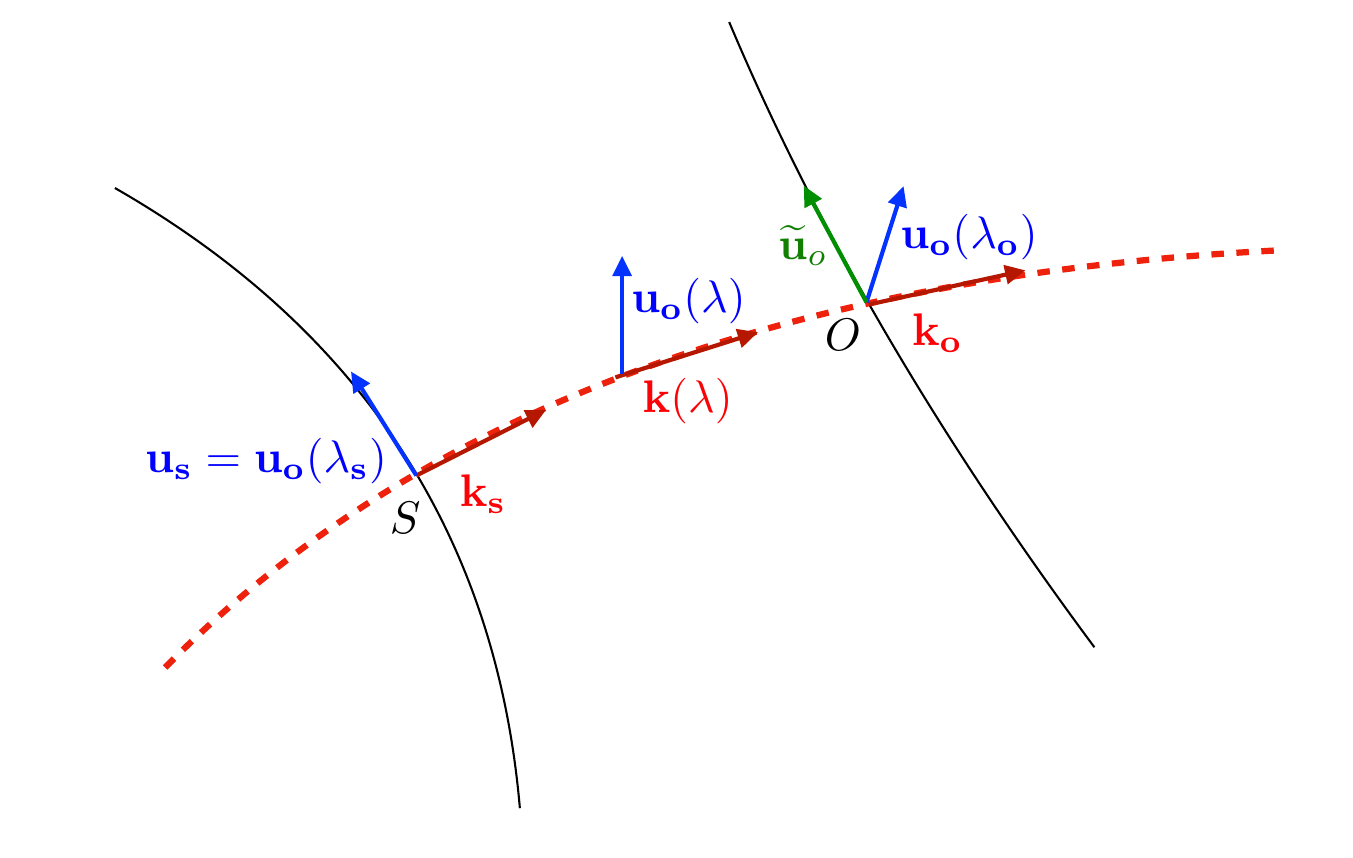}
\caption{\label{observer} The four-vector $u_o(\lambda)$ is
constructed by parallel-transporting $u_s$ from ${\cal S}$ to ${\cal O}$ along the null geodesic $\lambda$ connecting
them. The resulting vector $u_o(\lambda_o)\equiv u_o$ is by construction such that $z_{\text{ref}}=0$ and is called Synge source velocity. The spatial basis (not represented on the figure) at the source is also parallel transported along the geodesic, allowing to define a spatial basis for the reference observer (a Synge source spatial basis). The green vector represents the velocity of the observer $\widetilde{{\cal O}}$, which is related to ${\cal O}$ via a boost. The redshift of such an observer is $1+\tilde{z}=1/\mathcal{D}_o$. This figure is adapted from Fig.~1.5 of \cite{Fleury:2015hgz}. } 
\end{figure}

We construct a special reference observer for whom $z_{\text{ref}}=0.$  Let us start from the source tetrad ${\bm e}_a=({\bm u}, {\bm e}_i)$ that we take as a reference source frame (${\cal S}=\widetilde{\cal S}$),  
and let us parallel propagate it along the geodesic with the conditions
\be
k^\mu \nabla_\mu u^\nu = 0\,,\quad k^\mu \nabla_\mu e_i^\nu = 0\,.
\label{eq:parallel}
\ee
In Minkowski spacetime, a global tetrad satisfies $\nabla_\mu e_a^\nu = \partial_\mu e_a^\nu = 0$ and these conditions are automatically satisfied. However for a more general spacetime geometry, we cannot have $\nabla_\mu e_a^\nu =0$ everywhere, but we can require the milder conditions~\eqref{eq:parallel} for the geodesics emanating from a given source, as long as they do not intersect.\footnote{If geodesics intersect at the observer, then we receive several images (GW signals) of the source, but we can still define a tetrad by parallel propagation for each geodesic individually.}
This allows us to define the components of the GW everywhere during propagation. In particular we can decide to use such tetrad fields at the observer position as a reference observer frame ${\cal O}$. Eq.~\eqref{eq:parallel} combined with the geodesic equation~\eqref{geodesic} leads to $k^\mu \nabla_\mu (k_\nu u^\nu) = 0$, hence the energy is conserved along the geodesic, and therefore for this observer $\mathcal{O}$ we have $z_{\rm ref}=0$. A physical observer $\widetilde{{\cal O}}$ would in general be boosted with respect to this reference observer and from \eqref{JointDoppler} the redshift would then be $1+\tilde z=\Do^{-1}$. This construction corresponds to attributing the whole of the redshift to the local motion of the observer with respect to the reference observer. A schematic representation of the situation is depicted in Fig.~\ref{observer}.

This procedure was originally proposed by Synge~\cite{Synge1960}, and hence it was suggested in~\cite{Fleury:2015hgz} to name \emph{Synge source velocity} this reference observer velocity built from the source velocity. Since we also parallel transport the spatial basis we can also name \emph{Synge source tetrad} the whole tetrad built at the observer from the one of the source. The Synge source velocity is the best notion the true observer can have of the source velocity. 

With the same arguments, this constructions implies that $k^\mu \nabla_\mu (k_\nu e_i^\nu) = 0$ and from the decomposition~\eqref{kdecomposition} this implies that the components of the directions are constant during propagation, hence the reference source ${\cal S}$ and reference observer ${\cal O}$ agree on the direction of the GW, that is 
\be\label{nsnon}
n^s_i = n^o_i \equiv n_i\,.
\ee 
However the direction measured  by the true observer $\widetilde{{\cal O}}$ will in general be aberrated since $\widetilde{n^o_i} = R_{ij}^{o} n^j$.

\subsubsection{Synge observer and reference source}

Conversely we could start from the tetrad defined at the observer and use it as a reference observer (${\cal O}=\widetilde{\cal O}$) and parallel transport backward in time along the geodesic toward the source so as to defined a reference source frame ${\cal S}$. This would correspond to defining a reference source whose velocity is the most naturally associated to the true observer velocity, i.e.\ it would be the velocity obtained by parallel transporting the observer velocity backward in time along the geodesic. We  name the \emph{Synge observer tetrad} the tetrad built at the source by parallel transporting the tetrad of the observer.  By construction the Lorentz transformation between the Synge source tetrad and the observer tetrad (which are both defined at the observer position), is the inverse of the transformation relating the Synge observer tetrad and the physical source tetrad (which are both defined at the source position), hence in the latter case the redshift is given by $1+\tilde z = \Ds$. This alternative construction, which we do not use hereafter, corresponds to attributing the whole redshift to the local motion of the source. In that case the direction of emission for the true source $\widetilde{{\cal S}}$ is aberrated with respect to the direction of emission in the reference source frame ${\cal S}$ since $\widetilde{n^s_i} = R_{ij}^{s} n^j = R_{ij}^{o\, -1} n^j$.

\subsection{Observed waveform}
\label{sec:observedGW}

We now relate the wave emitted by a boosted source and received by the true (boosted) observer to the wave emitted by the reference source and observed by the reference observer. We proceed in two steps: we first consider a reference source and we consider how the signal it emits is received by the reference observer and the moving observer. Second, we consider a boosted source and the signal received from it.

\subsubsection{Signal from the reference source}

With a Synge source tetrad built as detailed in section~\ref{SecSyngeSource}, it is straightforward to express the components of the GW emitted by the source along the geodesic. Indeed, eq.~\eqref{conservation2} indicates that polarization is parallel transported, hence in such tetrad basis its components $\epsilon_{ij}$ are constant along the geodesic.\footnote{From the scaling of the amplitude~\eqref{ampl} we then deduce that $\chi \,h_{ij}$ is a constant along the geodesic.} We now consider a binary system of compact objects. Using for simplicity the quadrupole approximation, the components of the wave received by the reference observer in the direction $n_i$ (noted ${\cal O}(n_i)$) from the reference source are
\be\label{chihij1}
{h_{ij}}({\cal S} \to {\cal O}(n_i),\tau_o) = \frac{4 G}{c^4 \chi({\cal S} \to {\cal O})} \Lambda({{\bm n}})_{ij}^{\,\,kl} {T_{kl}}({\tau}_s(\tau_o))\,,
\ee
where ${T_{ij}}$ is the spatial part of the reference source energy momentum tensor. The function ${\tau}_s(\tau_o)$ is the relation between the proper time of the source and the one of the observer. The TT projector $\Lambda$ is defined for a given direction ${\bm n}$ by
\be\label{DefTTProj}
\Lambda({\bm n})_{ij}^{\,\,kl} \equiv \perp_i^k \perp_j^l - \frac{1}{2}\perp_{ij}\perp^{kl}\,,\qquad \perp_{ij} \equiv \delta_{ij} - n_i n_j \,.
\ee
Due to the special choice of reference observer ($z_{\rm ref}=0$), from~\eqref{dPhidtau3} we find that the relation between proper time of source and observer is simply given by $\tau_s = \tau_o+{\rm cst}$. 
The signal~\eqref{chihij1} can be written as
\be\label{hijBeij}
h_{ij}({\cal S} \to {\cal O}(n_i),\tau_o) = \frac{\Bsource\left({\bm n},T_{ij}({\tau}_s(\tau_o))\right)}{\chi({\cal S} \to {\cal O})} \epsilon_{ij}\,,
\ee
with the intrinsic amplitude and the polarization tensor defined as
\be\label{DefBepsij}
\Bsource\left({\bm n},T_{ij}\right) \equiv \frac{4G}{c^4} \sqrt{\Lambda({\bm n})_{ij}^{kl} T^{ij}T_{kl}}\,,\qquad \epsilon_{ij} = \frac{4G}{c^4}\frac{\Lambda({\bm n})_{ij}^{kl} T_{kl}}{\Bsource\left({\bm n},T_{ij}\right)}\,.
\ee
Note that this generalises eq.\,(\ref{amplitude}), where for illustration purposes, it was assumed that the GW was isotropically emitting radiation, i.e.\ that the amplitude $\mathcal{B}$ was independent of directions.

If we now want to determine the signal received by a true observer $\widetilde{{\cal O}}$ from this reference source, we need only to take into account the aberration of polarization~\eqref{Eq:guessepsilon} since from~\eqref{strainrule1b} the amplitude is conserved (the conservation of amplitude is also true for sources with anisotropic emission considered in this section, since a boost at the observer is independent on the properties of the signal emitted, i.e. it is a boost on received quantities). Then we get  
\be\label{hSOtohSOtilde}
h_{ij}({\cal S} \to \widetilde{\cal O}(\widetilde{n^o_i}),\tilde\tau_o)= R_i^{o\,k} R_j^{o\,l} h_{kl}({\cal S} \to {\cal O}(n_i),\tau_o(\tilde \tau_o))\,.
\ee
The notation $\widetilde{\cal O}(\widetilde{n^o_i})$ indicates that the signal is measured by the boosted observer $\widetilde{\cal O}$ in the aberrated direction $\widetilde{n^o_i}$. In practice, there can even be an additional rotation because there is no reason the true observer uses the same orientation as the one built with a simple boost from the reference observer (in other words, there is the residual freedom of a  3D rotation with respect to the previous construction).  The effect of redshift is only seen through the relation between the source proper time and the boosted observer proper time which becomes $\dd \tau_s/\dd \tilde \tau_o = \Do$ since $\dd \tau_o/\dd \tilde \tau_o = \Do$.

\subsubsection{Signal from boosted source}

If we consider a boosted source $\widetilde{{\cal S}}$ with associated tetrad $\tilde{e}_a^\mu$, then by parallel transporting this tetrad along the geodesic until the observer position, we define a boosted Synge source tetrad at the observer $\widetilde{\cal O}$,  whose boost relative to the initial Synge source tetrad ${\cal O}$ is the same as the one that relates the boosted source $\widetilde{\cal S}$ to the reference source ${\cal S}$ frames (at the source position). Hence directions are related by~\eqref{nsnon} and 
\be
\widetilde{n^o_i} = R_i^{\,j} n_j\,,\qquad \widetilde{n^s_i} = R_i^{\,j} n_j\,, \qquad R_{ij} \equiv R^s_{ij}(v^s_k,n_k)\,.
\ee
The signal of the boosted source seen by the boosted Synge observer (i.e.\ the observer defined by the Synge boosted source tetrad) is
\be\label{tildeev}
h_{ij}(\widetilde{\cal S} \to \widetilde{\cal O}(\widetilde{n^o_i}),\tilde\tau_o) = \frac{4 G}{c^4 \chi(\widetilde{\cal S} \to {\widetilde{\cal O}})} \Lambda({\widetilde{\bm n}})_{ij}^{\,\,kl} {T_{kl}}(\tilde{\tau}_s(\tilde\tau_o))\,,
\ee
that is the same expression as~\eqref{chihij1}, but with tildes everywhere.\footnote{The components of the energy momentum tensor of the boosted source in the boosted basis are the same as the reference energy momentum tensor in the reference basis. Indeed the reference energy momentum tensor is $T^{\mu\nu}=T^{ab} e_a^\mu e_b^\nu$ while the boosted one is $\widetilde{T}^{\mu\nu}=T^{ab} \tilde{e}_a^\mu \tilde{e}_b^\nu$.} 
 
Then we know that the wave amplitude seen by the two observers  is the same [eq.~\eqref{strainrule1b}], but polarization is aberrated according to~\eqref{Eq:guessepsilon}, hence 
\be\label{tildehij}
h_{ij}(\widetilde{\cal S} \to \widetilde{\cal O}(\widetilde{n^o_i}),\tilde\tau_o) = R_i^{\,k} R_j^{\,l} {h}_{kl}(\widetilde{\cal S} \to {\cal O}({n_i}),\tau_o)\,.
\ee
Using \eqref{MegaTchi} and the property of the TT projectors
\be\label{RRRRLambda}
R_i^{\,\,k} R_j^{\,\,l} R_{\,\,p}^{r} R_{\,\,q}^{s}\Lambda({\bm n})_{kl}^{\,\,pq}  = \Lambda(\tilde{\bm n})_{ij}^{\,\,rs}\,,
\ee
the signal of the boosted source seen by the reference observer is found by inversion of~\eqref{tildehij} and is given by 
\begin{equation}\label{MasterFormula}
{h}_{ij}(\widetilde{\cal S} \to {\cal O}(n_i),\tau_o) = \Ds^{-1} \frac{4 G}{c^4 \chi({\cal S} \to {\cal O})} \Lambda({{\bm n}})_{ij}^{\,\,kl} R_k^{-1\,p} R_l^{-1\,q }{T_{pq}}(\tilde{\tau}_s(\tau_o))\, .
\end{equation}
The relation between the boosted source proper time and the observer proper time is $\dd \tilde \tau_s/\dd \tau_o = \Ds^{-1}$.
To find the relation between amplitudes of (\ref{hijBeij}) and (\ref{MasterFormula}),  we need to use that for a projector $\Lambda({\bm n})_{ij}^{kl}\Lambda({\bm n})_{kl}^{pq} = \Lambda({\bm n})_{ij}^{pq}$ combined with property~\eqref{RRRRLambda} and this gives 
\be\label{newrule}
{H}(\widetilde{\cal S} \to {\cal O}(n_i),\tau_o) = \Ds^{-1}  H({\cal S} \to {\cal O}(n_i),\tau_o) \frac{\Bsource\left(\tilde{\bm n},{T_{ij}}(\tilde{\tau}_s(\tau_o))\right)}{\Bsource\left({\bm n},{T_{ij}}({\tau}_s(\tau_o))\right)}\,,
\ee
which generalizes~\eqref{strainrule1} for anisotropic sources. 

The ratio of $\mathcal{B}$ functions on the right hand side of (\ref{newrule})  reflects the fact that, due to the source velocity, we see a \emph{different side} of the source: we receive indeed the GW radiation emitted in direction $-\tilde{\bm n}$ instead of the radiation emitted in direction $-{\bm n}$. If the source emits isotropically, this does not change the observed amplitude, as is apparent from eq.~\eqref{strainrule1}. However, if the source emits anisotropically, then seeing a different side does change the observed amplitude. 

Finally if we want the signal of a true observer with respect to this boosted source, then we must add the aberration rotation due to the observer motion as in \eqref{hSOtohSOtilde}, hence we combine~\eqref{MasterFormula} with\footnote{This has formally the same structure as (\ref{tildehij}), with the difference that here we are using the matrix $R^o_{ij}$ and not $R_{ij}=R^s_{ij}$.}
\be\label{hStildeOtohStildeOtilde}
h_{ij}(\widetilde{\cal S} \to \widetilde{\cal O}(\widetilde{n^o_i}),\tilde\tau_o)= R_i^{o\,k} R_j^{o\,l} h_{kl}(\widetilde{\cal S} \to {\cal O}(n_i),\tau_o(\tilde \tau_o))\,.
\ee
This does not affect the transformation~\eqref{newrule} of the magnitude, but the relation between proper times now involves the joint Doppler factor as it becomes $\dd \tilde \tau_s/\dd \tilde \tau_o = {\cal D}$ since $\dd \tau_o/\dd \tilde \tau_o = \Do$.

\subsection{Degeneracy between sources}
\label{sec:degeneracies}

Eq.~\eqref{newrule} together with eqs.~\eqref{Tdtdt} and~\eqref{eq:pol_trans} tell us how the GW amplitude, frequency and polarization are affected by the source velocity. 
In this section we show that we can build a class of equivalent sources sharing the same observed signal. This has profound consequences, since it means that the source velocity cannot be inferred from the measurement of the GW waveform.

We start from the reference source and observer of section~\ref{Synge} (the reference observer velocity is the Synge reference source velocity such that $z_{\rm ref}=0$), and build  a class of sources located at the same position as the reference source, emitting a signal that is totally degenerate with the signal emitted by the reference source when observed by the reference observer. 
It is clear that boosted sources $\widetilde{\cal S}$ from which the signal received by the observer is the same, must have the same redshift as the reference source, hence they must have a Doppler shift $\Ds=1$.\footnote{Note that we could have $\mathcal{D}_s\neq 1$ and still have an equivalent source by placing it at a larger or smaller distance. However here we assume that the determination of distance is possible thanks to the knowledge of the host galaxy, hence we restrict to the case $\mathcal{D}_s=1$.} In that case the the propagation distances are the same, $\chi(\widetilde{{\cal S}} \to {\cal O}) = \chi({\cal S} \to {\cal O})$, meaning that the factor $\mathcal{D}_s^{-1}$ in~\eqref{newrule} is unit, and the time evolution is not affected since $\tilde \tau_s = \tau_s$.
From this condition and the definition of the Doppler factor \eqref{alphadef} we find that for a given boost velocity with norm $\beta$, the angle $\theta$ between the velocity direction $\hat v^i = v^i/\beta$ and the observing direction needs to satisfy
\be\label{thetaDone}
\cos \theta = n_i \hat v^i = \frac{1-\sqrt{1-\beta^2}}{\beta}\,.
\ee
The angle $\theta$ ranges from $\pi/2$ to $0$ when $\beta$ goes from $0$ to its maximum value $1$. It follows that the condition $\Ds=1$ can only be obtained for sources moving away from the observing direction (in the sense that $n_i \hat v^i \geq 0$). Eq.~\eqref{thetaDone} tells us that there is a family of moving sources, with velocity $\beta$ going from 0 to 1, that have exactly the same Doppler factor $\Ds=1$ as the reference source. The direction of the velocity for these sources must be very specific and obey~\eqref{thetaDone}, in order to cancel the effect of the velocity in $\Ds$. 

From eqs.~\eqref{newrule} we see that having $\mathcal{D}_s=1$ is however not enough to see the same observed signal. The ratio of $\mathcal{B}$ functions on the right hand side of (\ref{newrule}) also need to be equal to 1. This cannot be achieved by an appropriate choice of velocities, but it can be done by considering boosted sources that are rotated with respect to the reference one. Indeed if we start with a source with spatial energy momentum tensor
\be\label{DefTrot}
T^{\rm rot}_{ij} \equiv R_i^{\,\,k} R_{j}^{\,\,l} T_{kl}\,,
\ee
from eqs.~(\ref{DefBepsij}) and \eqref{RRRRLambda} we obtain
\be\label{Brotiota}
\Bsource(\tilde{{\bm n}},T_{ij}^{\rm rot}) \equiv \frac{4G}{c^4} \sqrt{\Lambda(\tilde{\bm n})_{ij}^{kl} T^{ij}_{\rm rot} T^{\rm rot}_{kl}} = \Bsource({\bm n},T_{ij})\, , 
\ee
and therefore
\be\label{HrotH}
H(\widetilde{\cal S}^{\rm rot} \to {\cal O}) = \Ds^{-1} H({\cal S} \to {\cal O})=H({\cal S} \to {\cal O})\, ,
\ee
for $\Ds=1$. The rotation of the source defined in~\eqref{DefTrot} ensures us that, after aberration, we see the same part of the source that we would see if the source was not moving. A good analogy is a lighthouse: if the source $\mathcal{S}$ is emitting like a lighthouse (e.g.\ pulsar like), then if the signal it emits is received by $\mathcal{O}$, the signal from the boosted source $\tilde{\mathcal{S}}$ cannot be received because of aberration, unless the source $\tilde{\mathcal{S}}$ is additionally rotated with respect to $\mathcal{S}$ to compensate for the effect of aberration. 

Finally, from~\eqref{Eq:guessepsilon}, we see that the rotation applied to the energy-momentum tensor of the boosted source does exactly compensate for the rotation of the polarization, such that the boosted and rotated source has exactly the same polarization as the reference source.

To summarize, starting from a reference source, we can build boosted sources which produce the same observed signal and which have arbitrary large boost magnitudes $\beta$. Indeed, for any  $\beta$ we can find the orientation of the boost velocity needed to keep $\mathcal{D}_s=1$ from~\eqref{thetaDone} and therefore the aberration rotation $R_i^{\,\,j}(v_k,n_k)$. This rotation can then be used to rotate the boosted source orientation as in~\eqref{DefTrot}, so as to erase {\it i)} the effect of aberration on the observed polarization and {\it ii)} the effet of aberration on the magnitude induced by the directional dependence of $\Bsource$. This is due to the fact that, in practice, we cannot disentangle the rotation of the emitted direction (and polarization) due to aberration and the one due to a different orientation of the source itself. 

The fact that such a class of sources (with exactly the same observed signal) exists, means that it is not possible to measure the source velocity from the GW signal. This was already pointed out in~\cite{Bonvin:2022mkw}, looking at the effect on polarization at linear order in velocity.\footnote{Note that \cite{Bonvin:2022mkw} corrects some previous statements of \cite{PhysRevD.104.123025, PhysRevD.100.063012}, where the authors erroneously claim that the effect of aberration gives more than a simple spin phase $\delta$ on the polarization helicity components, and that the velocity can therefore be observed.} 
This degeneracy is general and valid beyond the quadrupole formula, and also for electromagnetic waves (in the eikonal limit), as long as we can only see one direction of emission. This excludes cases in which we can observe two different directions of emission, e.g.\ via a lensed signal which arrives from another direction. 

As an observational consequence, if we know $\Ds$, we cannot attribute it to a unique source velocity. 
The argument derived above for $\Ds=1$ can indeed be repeated for any value of $\Ds$, to show that also in this case there exist a whole family of sources with different velocities but the same observed signal. One possibility is to resort to the simplest explanation, which consists in attributing the redshift to a radial boost, in order to select one source in the family of equivalent ones, namely the source for which the transformation between the Synge source velocity and the true observer velocity is a boost in the direction of observation. This arbitrary choice would bias the measurement of the orientation of the source, since it is degenerate with the velocity direction (angle between the velocity and observation directions). It would however not bias the determination of the other source parameters (e.g.\ masses and spins) and the cosmological parameters, since those are only degenerate with $\Ds$ that we assume here to be known.

In practice however, we generally do not know $\Ds$.  Even if we can measure the redshift of the source (e.g.\ from an electromagnetic counterpart) we do not know which part of the redshift is due to the Hubble flow, and which part is due to the source peculiar velocity. In this case, all parameters of the source and the cosmological parameters are biased, since the amplitude of the wave and the frequency are modified by this unknown $\Ds$.

\section{Waveform in the cosmological context}
\label{cosmo}

We consider a cosmological context in which both the source and the observer move with respect to a reference frame, usually identified with the CMB rest frame, i.e.\ the reference source and the reference observer are comoving observers of the Friedmann-Lema\^itre-Robertson-Walker cosmology characterized by a time evolving scale factor $a$. Contrary to the previous sections, in the cosmological context the reference source and the reference observer are not at rest with respect to each other: they move away from each other due to the expansion of the universe. We start therefore by constructing the reference frame and the reference signal, accounting for this relative velocity, that is called the Hubble flow. And we then extend the calculation to account for peculiar velocities of the source and the observer, on top of the Hubble flow.

\subsection{Reference frame}\label{ref}


We choose a reference source which is comoving with the Hubble flow, and we build from it the Synge source. By construction, the redshift seen by the observer associated to the Synge source is $z_{\text{ref}}=0$.  One can then relate the Synge frame at the observer to the frame of the observer comoving with the Hubble flow (that we want to use as a reference).  The two frames are related via a radial boost such that $\bar{\mathcal{D}}_o=1/(1+\bar{z})=a_s/a_o$. Note that here and in the following, we denote with a bar quantities related to background Hubble frame. Using the expression of the Doppler shift for radial velocities~\eqref{SimpleD}, this determines  the boost velocity needed to go from a Synge source frame to a frame comoving with the Hubble flow, which is $\bar{v}_i = - \bar{v}_H \bar{n}_i$ with\footnote{$\bar{v}_H$ can be interpreted as a recession velocity even for large redshifts, since for small redshifts $a_s = a_0 - \delta a$ and we recover the usual relation between recession velocity and distance $\bar{v}_H\simeq \delta a/a_o \simeq H_o \delta t \simeq H_o \bar\chi$.}
\be\label{DefvH}
\bar{v}_H = \frac{a_o^2- a_s^2}{a_o^2 + a_s^2}\,.
\ee
Since the boost is radial, $R_{ij}(\bar{v}_i,\bar{n}_i)=\delta_{ij}$, and there is no aberration due to the universe's expansion. The  relation between the source proper time and the new reference observer, which is comoving with the Hubble flow, is however affected by the expansion 
\be\label{dtaudtaucosmo}
\frac{\dd \bar \tau_s}{\dd \bar \tau_o} = \frac{1}{(1+ \bar z)}\,.
\ee

Alternatively, one could require that the timelike vector of the tetrad $u^\mu$ always matches the velocity of the Hubble flow $\bar u^\mu$. However the latter velocity is not parallel transported and satisfies instead $k^\mu \nabla_\mu \bar u^\nu = \bar{n}^\nu \dd \ln a / \dd \lambda$. Nonetheless, we could still ensure that the components of the direction vector $\bar{n}_i$ are conserved along a given geodesic by defining the spatial vectors of the tetrad with  
\be\label{MagiceTransport}
k^\mu \nabla_\mu \bar{e}_i^\nu = \frac{\dd \ln a}{\dd \lambda} \bar{n}_i \bar{u}^\nu \,,
\ee
as it implies $k^\mu \nabla_\mu \bar n_i = k^\mu \nabla_\mu(-{\bm k} \cdot {\bm{\bar{e}}}_i / {\bm k} \cdot {\bm{\bar{u}}})=0$. This construction amounts to allowing infinitesimal boosts in the radial direction (the infinitesimal velocity components of such boosts being $(\dd \ln a / \dd \lambda) \bar{n}_i$) so as to always select the desired Hubble flow velocity along a given geodesic. Since boosts in the radial direction to not induce aberration, the components of the direction vector in this basis must remain constant. The final result is the same as when considering a pure Synge source tetrad (that is parallel transported with~\eqref{eq:parallel}) to which we eventually add a radial boost with velocity~\eqref{DefvH}. Either the radial boosts are added gradually along the geodesic, or a single boost is added at the very end of a pure parallel transport. The construction~\eqref{MagiceTransport} is possible because the Friedmann-Lemaître-Robertson-Walker (FLRW) geometry is conformally related to a Minkowski geometry by a scale factor.

\subsection{Reference signal}

In practice, the usual spherical basis of the spatial section of the FLRW geometry (even if spatial sections are curved) when the coordinates system is centered on the source satisfies the requirement~\eqref{MagiceTransport}. Hence, the results expressed in a Synge source frame can be transposed trivially to the FLRW case with such choice of basis, as one needs only to take into account the modified relation~\eqref{dtaudtaucosmo} between proper times.

Let us consider a binary system of compact objects in the Newtonian approximation with masses $M_1$ and $M_2$. In the source frame, we define a spherical basis $(\bar{e}_r^i = \bar{n}^i$, $\bar{e}_\theta^i$, $\bar{e}_\phi^i)$, with the azimuthal direction aligned with the normal to the binary, that is $N^i = \bar{e}_z^i$. The time varying part of the spatial components of the energy momentum tensor is 
\be\label{WhyNotPuttingTijAsWell}
T^{ij} = M_c (G M_c  \pi f_s)^{2/3}\left[\cos (2\Phi_{\rm orb}) \bar{e}_x^i \bar{e}_x^j-\sin (2\Phi_{\rm orb}) \bar{e}_y^i \bar{e}_y^j\right]\,,
\ee
where $M_c \equiv (M_1 M_2)^{3/5}/(M_1+M_2)^{1/5}$ is the chirp mass and $\Phi_{\rm orb}$ is  the phase of the circular orbital motion. From the components $h_{ij}$ of the signal we define polarization components $h_{(+)}$ and $h_{(\times)}$ as in~\eqref{Defpluscross}, and from the quadrupole formula~\eqref{chihij1} with~\eqref{WhyNotPuttingTijAsWell}, these components for a source and observer comoving with the Hubble flow, take the form~\cite{Maggiore:1900zz}
\begin{subequations}\label{0PN}
\begin{align}
\bar{h}_{(+)}&=\frac{4G {M}_c}{\bar{\chi}}(G {M}_c\pi f_s)^{2/3} \frac{1+\cos^2\bar{\iota}}{2}\cos(\Phi)=\frac{4G\bar{\mathcal{M}}_c}{\bar{D}_L}(G \bar{\cal M}_c\pi \bar{f}_o)^{2/3} \frac{1+\cos^2\bar{\iota}}{2}\cos(\Phi)\,,\\
\bar{h}_{(\times)}&=\frac{4 G M_c}{\bar{\chi}}( G {M}_c\pi f_s)^{2/3}\cos\bar{\iota}\sin(\Phi)=\frac{4 G \bar{\mathcal{M}}_c}{\bar{D}_L}( G \bar{{\cal M}}_c\pi \bar{f}_o)^{2/3}\cos\bar{\iota}\sin(\Phi)\,,
\end{align}
\end{subequations}
where a bar denotes quantities defined in the frame comoving with the Hubble flow. Here, $\Phi= 2 \Phi_{\rm orb}  = 2\pi \int \dd \bar{\tau}_s f_s $ is the phase of the wave whose evolution needs to be related to the observer proper time with~\eqref{dtaudtaucosmo}. The redshifted chirp mass and the observed frequency are related to intrinsic quantities at the source via
\be\label{DefcalMc}
\bar{\mathcal{M}_c}(\bar{z})\equiv(1+\bar{z})M_c\,,\qquad \bar{f}_o=f_s/(1+\bar{z})\,,
\ee
so that $M_c f_s = \bar{{\cal M}}_c \bar{f}_o$. The inclination angle $\bar{\iota}$ is the angle between the normal to the binary plane ${\bm N}$ and the direction of emission $\bar{{\bm n}}$, that is 
\be\label{Defbariota}
\cos \bar{\iota} \equiv N^i \bar n_i\,.
\ee

\subsection{Perturbations of redshift and luminosity distance}

 When both the source and the observer move with respect to the background Hubble flow the redshift is denoted as $z$. We consider cosmological peculiar velocities, that are typically small, such that we can work at linear order. From \eqref{JointDoppler}, and linearizing the Doppler factors we obtain
 \be\label{ztozbar}
 (1+z)=(1+\bar{z})[1-(\bv_o-\bv_s)\cdot {\bm{n}}]\,.
 \ee
The luminosity distance kinematic fluctuations have been derived in~\cite{Bonvin:2005ps}. They can also easily be calculated from~\eqref{MegaTchi} using that $D_L({\cal S} \to {\cal O}) = \bar D_L$ and linearizing the Doppler shift factors. It is also convenient to realize that $\bar{\chi}\equiv \chi(\mathcal{S}\rightarrow \mathcal{O})$ is the standard metric distance in a FL universe, which in a flat universe is related to conformal time by $\bar{\chi}(\eta)=\int_{\eta}^{\eta_0} d\eta'$, hence the simplest derivation is 
\be
\label{eq:DLcosmo}
D_L=(1+z)\chi(\tilde{\mathcal{S}}\rightarrow \tilde{\mathcal{O}})=(1+z)\chi(\tilde{\mathcal{S}}\rightarrow \mathcal{O})=(1+z)\mathcal{D}_s\,\chi(\mathcal{S}\rightarrow \mathcal{O})=(1+z)(1+{\bm{n}}\cdot\bv_s)\bar{\chi}\, .
\ee
From this, we obtain for the luminosity distance at fixed conformal time
\be\label{DLeta}
D_L(\eta, \bn)=\bar{D}_L(\eta)\left[1-{\bm{n}}\cdot (\bv_o-2\bv_s)\right]\,.
\ee
The luminosity distance at fixed redshift is then
\be\label{DLz}
D_L(z, {\bm{n}})=\bar{D}_L(z)\left[1+{\bm{n}}\cdot \bv_s+(\bv_o-\bv_s)\cdot {\bm{n}} \frac{1}{(\eta_o-\eta)\mathcal{H}}\right]\,,
\ee
which was obtained from the relation
\be
D_L(\eta, {\bm{n}})=D_L(z, {\bm{n}})-\frac{\dd}{\dd\bar{z}}\bar{D}_L(\bar{z})\delta z\,,
\ee
with 
\be
\frac{\dd}{\dd\bar{z}}\bar{D}_L(\bar{z})=\frac{\bar{D}_L}{(1+\bar{z})}+\frac{1}{\mathcal{H}}\,,\qquad
\delta z=(1+\bar{z})(\bv_s-\bv_o)\cdot {\bm{n}}\,.
\ee
As already discussed in section~\ref{sec:interpretation} the luminosity distance is not symmetric in the source and observer velocity, i.e.\ it does not depend only on the difference $\bv_o-\bv_s$. From eq.~\eqref{eq:DLcosmo} we see that this asymmetry is directly linked to the fact that the distance $\chi(\tilde{\mathcal{S}}\rightarrow \tilde{\mathcal{O}})$ is only affected by the source velocity. We emphasize that this is a physical effect, related to the fact that distances are not invariant under a boost. As explained in section~\ref{sec:interpretation} this does not lead to any inconsistency: it simply reflects the fact that a moving source emitting a signal at the same time as a reference source has not the same impact on distances than a moving observer receiving a signal at the same time as a reference observer.

\subsection{Signal from boosted source to boosted observer}

Let us write the waveform in the Newtonian approximation for a boosted source and observer, as done in  (\ref{0PN}) for a source and observer comoving with the Hubble flow.  We first assume that the source does not have any electromagnetic counterpart associated, so we do not have access to redshift information. We also neglect the spin-phase effect on polarizations, since it can be reabsorbed in a redefinition of the polarization basis at the observer. We must however take into account which side of the source is seen by the observer, since this is modified by aberration. Indeed in the source frame, the direction of emission is aberrated and given by $n_i = R^s_{ij} \bar n^j$, but the components $N^i$ of the normal to the boosted binary are unchanged,\footnote{For the reference source, the normal to the binary plane is ${\bm N} = N^i {\bm e}_i$, but for the boosted source the normal vector becomes $\widetilde{\bm N} = N^i \tilde{\bm e}_i$, hence the components are the same in the respective basis.} hence the side seen is characterized by
 \be
\cos \iota \equiv N^i n_i = N^i R^s_{ij} \bar n^j\,,
 \ee
and in general $\iota \neq \bar{\iota}$. This geometrical statement is equivalent to the appearance of the $R^{-1}_{ij}$ rotations in~\eqref{MasterFormula}, meaning that we do not see the same side as for the signal emitted by a reference source. In other words,  we receive a signal which is exactly the one of a reference source that one would have additionally rotated such that its normal vector is $N^{\rm rot}_i \equiv N^j R^s_{ji} = R^{s\,-1}_{ij} N^j$, since in that case $\cos \iota = \cos \bar{\iota}^{\,\rm rot} \equiv N^{\rm rot}_i \bar n^i$ by construction.
 
The signal received by a moving observer from a moving source is formally the same as~\eqref{0PN} but without bars, and using~\eqref{eq:DLcosmo} we relate it to the signal of the reference situation by
\begin{subequations}
\begin{align}
h_{(+)}
&=\frac{(1+z)}{D_L}4(G{M}_c)^{5/3}(\pi f_s)^{2/3} \frac{1+\cos^2\iota}{2}\cos(\Phi)=(1-\bv_s\cdot {\bm{n}})\bar{h}_{(+)}(\cos\iota)\, ,\label{eq:hfixedetaplus}\\
h_{(\times)}
&=\frac{(1+z)}{D_L}4(G{M}_c)^{5/3}(\pi f_s)^{2/3} \cos\iota\sin(\Phi)=(1-\bv_s\cdot {\bm{n}})\bar{h}_{(\times)}(\cos\iota)\, ,
\label{eq:hfixedetacross}
\end{align}
\end{subequations}
up to a spin phase which mixes these components as in~\eqref{spinphasecossineffect}. It is implied that the relations between proper times and frequencies become
\be\label{DefcalMc2}
\frac{f_s}{f_o}=\frac{\dd \tau_o}{\dd \tau_s}=1+z\,,
\ee
with the redshift $z$ related to the reference redshift $\bar{z}$ through~\eqref{ztozbar}.

The observed orientation of the binary is directly affected by the source velocity: we do not see the same side of the source that one would observe in the absence of aberration. On the contrary, the angle $\iota$ does not depend on the observer velocity. The observer velocity affects the direction from which we receive the signal, but this has no impact on $\cos\iota$, since $\cos\iota$ is the angle between the orientation of the binary and the direction of the \emph{emitted} signal, that are both observer-independent. The angle $\iota$ is the only geometrical information we can extract from the polarization. Since for a radial source velocity, the aberration rotation is simply $R^s_{ij} = \delta_{ij}$, only transverse velocities (at linear order) can modify the side of the source which is seen, as was already found in eq.~(58) of~\cite{Bonvin:2022mkw}.

Finally, we see that the wave amplitude is affected only by the source velocity. The prefactor $(1-\bv_s\cdot {\bm{n}})$ is nothing but the factor $\Ds^{-1}$ which appears in~\eqref{MasterFormula}. The observer velocity generates aberration but it does not affect the amplitude. This is in line with what we found in the general case: the observer motion with respect to the reference observer (that in our case here is comoving with the Hubble flow) does not affect the strain. From~\eqref{eq:hfixedetaplus} and~\eqref{eq:hfixedetacross} we see that a source moving towards the observer (with $\bv_s\cdot {\bm{n}}<0$) has a strain which is amplified with respect to the reference source. This is related to the fact that such a moving source is closer to the observer at the time of reception of the GW than the reference source: $\chi(\tilde{\mathcal{S}}\rightarrow \tilde{\mathcal{O}})<\chi(\mathcal{S}\rightarrow \tilde{\mathcal{O}})$. As a consequence, its amplitude is less diluted than the one of the reference source.

Let us now assume that we also have an electromagnetic counterpart, hence we have a measurement of the redshift, and we compute the GW signal for sources that are at the same redshift (instead of the same conformal time as in eqs.~\eqref{eq:hfixedetaplus} and~\eqref{eq:hfixedetacross}). In this case, we need only to take into account perturbations to the luminosity distance (at fixed redshift) and we obtain 
\begin{align}
\label{eq:hfixedz}
h_{(+)}(z)
&=\left[1-{\bm{n}}\cdot \bv_s-(\bv_o-\bv_s)\cdot {\bm{n}} \frac{1}{(\eta_o-\eta_s)\mathcal{H}_s}\right]\bar{h}_{(+)}(z,\cos{\iota})\,,
\end{align}
where we have explicitly indicated that quantities are computed at fixed redshift. The same expression holds for $h_{(\times)}$. In this case, we see that the strain is affected by both the source and the observer velocity. This is simply due to the fact that two different observers (one moving and one at rest with respect to the Hubble flow) that are seeing the same source at the same redshift cannot be both at the same luminosity distance from the source. Hence they do not measure the same strain. This situation is therefore different from situation B of section~\ref{boostwave} where we compared two observers that are at the same position in space-time when they receive the signal. Here instead we compare two observers that see the source at the same redshift, and are therefore at different space-time positions. 

\subsection{Waveform and parameter extraction}

Let us now move to the impact on the waveform, i.e.\ the evolution of the frequency with proper time. The total energy of a circular binary in the Newtonian approximation is  ${\cal E} = -1/2(G \pi f_s M_c)^{2/3} M_c$. The energy flux per area is proportional to the energy density, hence knowing the energy density of GW from eq.~\eqref{rho}, we deduce that the total energy flux emitted through GW is given by $F = (64 \pi G)^{-1}\int \dd \chi^2 H^2 (2\pi f_s)^2 \dd \Omega_s$, which from the quadrupole formula components~\eqref{0PN} yields $F = 32/(5 G) (G M_c \pi f_s)^{10/3}$. From the energy balance $\dd {\cal E}/\dd \tau_s = - F$ we get the evolution of the frequency (hence at lowest order in post-Newtonian expansion)
\be\label{MagicChirp}
\frac{\dd f_s}{\dd \tau_s}= \frac{96 \pi}{5} f_s^2 \left(G M_c \pi f_s\right)^{5/3}\quad \Rightarrow \quad \frac{\dd f_o}{\dd \tau_o}= \frac{96 \pi}{5} f_o^2 \left(G {\cal M}_c \pi f_o\right)^{5/3}\,,
\ee 
where we used in the last step that frequencies and times are redshifted as in~\eqref{DefcalMc2} and we defined the redshifted chirp mass by ${\cal M}_c \equiv (1+z) M_z$. As detailed e.g.\ in \cite{Bonvin:2016qxr}, the last relation depicts how the frequency evolves for the observer and it implies that if the redshift does not vary over the observation time we can only deduce the redshifted chirp mass ${\cal M}_c$, and not the chirp mass $M_c$ directly, hence introducing a bias in its reconstruction.\footnote{If the redshift varies over the observation time, the functional dependence of $f_o$ with the proper time of the observer changes, for instance due to the evolution of the source scale factor, or due to peculiar acceleration. It has been shown that the effect of peculiar accelerations can be observable on the chirp in some astrophysical scenarios \cite{Bonvin:2016qxr, Tamanini:2019usx}.} This conclusion holds even if one considers post-Newtonian corrections since the last factor in the rhs of~\eqref{MagicChirp} is replaced by a more complicated function of $(G M_c \pi f_s)=(G {\cal M}_c \pi f_o)$ and of the mass ratio. It follows that observationally, we can only measure redshifted individual masses. 

Consequently, since the amplitude of the signal depends on the prefactor $M_c/\chi={\cal M}_c/D_L$, we can only infer, from the amplitude of the signal received, $D_L$ and not $\chi$. For this reason it is often stated that GW decay like the inverse of the luminosity distance. This is not true strictly speaking since from~\eqref{amplitude} the amplitude decays as $1/\chi = (1+z)/D_L$. What is meant is that when extracting the source parameters, we can treat the wave as a decreasing function of $D_L$ for a fixed redshifted chirp mass deduced from the frequency evolution.

\section{Discussion and conclusions}\label{conclusions}

In this article we have derived in detail how a gravitational wave emitted by a given source is affected by the velocities of the source and of the observer (with respect to some reference source and reference observer, that in a cosmological context we consider to be comoving with the Hubble flow). Our treatment is valid in the context of general relativity for any velocity, including relativistic ones. We have calculated the impact of velocities on all relevant quantities, namely the frequency, the amplitude, the polarization tensor and the energy density of the wave. The transformation properties of all quantities are summarized in table~\ref{Table1}.

\begin{table}[!htb]
\begin{center}
\begin{tabular}{ cccc } 
 \toprule
 & boosted source  & boosted observer & equations \\
 \midrule
 $E_o = 2\pi f_o$ & $1$ & $\Do$ & \eqref{DefDoDs}\\
 \midrule
 $E_s = 2\pi f_s$ & $\Ds$ & $1$& \eqref{DefDoDs}\\
 \midrule
 $1+z=\frac{E_s}{E_o} = \frac{\dd \tau_o}{\dd \tau_s}$ & $\Ds$ & $\Do^{-1}$ & \eqref{ztildeDsDo}, \eqref{Tdtdt}\\
 \midrule
 $\chi$ & $\Ds$ & $1$ & \eqref{eq:chis}, \eqref{chiunchanged}, \eqref{MegaTchi} \\
 \midrule
 $D_A$ & $1$ & $\Do$ &  \eqref{TDAA}, \eqref{TDAB}, \eqref{MegaTchi} \\
 \midrule
 $D_L$ & $\Ds^2$ & $\Do^{-1}$ & \eqref{TDLA}, \eqref{TDLB}, \eqref{MegaTchi} \\
 \midrule
 $H$ & $\Ds^{-1}$ & $1$ & \eqref{strainrule1}, \eqref{strainrule1b}\\
 \midrule
 $\dd \Omega_s$ & $\Ds^{-2}$ & $1$ & \eqref{TransdOmega2}\\
 \midrule
 $\dd \Omega_o$ & $1$ & $\Do^{-2}$ & \eqref{TransdOmega2}\\
 \midrule
 $\rho_{\rm GW}$ & $\Ds^{-4}$ & $\Do^2$ & \eqref{rhoT1}\\[1mm]
 \midrule
 $\displaystyle \frac{\dd\rho_{\rm GW}}{\dd\Omega_o \dd\log E_o}$ & $\Ds^{-4}$ & $\Do^4$ & \eqref{Tspecific1}\\[3mm]
 \midrule
 $\displaystyle \frac{\dd\rho_{\rm GW}}{\dd\Omega_o}$ & $\Ds^{-4}$ & $\Do^4$ &  \eqref{Tspecific2}\\[3mm]
 \midrule
 $\displaystyle \frac{\dd\rho_{\rm GW}}{\dd\Omega_o \dd E_o}$ & $\Ds^{-3}$ & $\Do^3$ & \eqref{Tspecific3}\\[3mm]
 \bottomrule
\end{tabular}
\caption{Doppler transformation factors (defined in section~\ref{SecRedshift}) with respect to the reference situation, for isotropic sources. The quantity of each line in a given situation is obtained from the factor written multiplied by the corresponding quantity in the reference situation. For instance we get ${H}(\widetilde{\mathcal{S}}\rightarrow {\mathcal{O}})=\Ds^{-1} H(\mathcal{S}\rightarrow \mathcal{O})$ and ${H}({\mathcal{S}}\rightarrow \widetilde{\mathcal{O}})= H(\mathcal{S}\rightarrow \mathcal{O})$ as obtained in eqs.~\eqref{strainrule1} and \eqref{strainrule1b}.}\label{Table1}
\end{center}
\end{table}

We first studied the case of a source emitting isotropically. In this case we have shown that the amplitude of the wave is independent of the observer velocity, while it gets amplified with a Doppler factor if the source moves (with respect to a reference source). In other words, while in the aberration context source and observer velocities play a symmetrical role, this is not the case for the amplitude (not even in Minkowski spacetime). The reason is that the strain amplitude scales as $1/\chi$ where $\chi=\sqrt{\dd A_o/\dd\Omega_s}$ and $\dd\Omega_s$ corresponds to the angle of the emitted bundle that is received by the observer and $\dd A_o$ is the surface of the bundle at the observer. This quantity is the same for all observers related by a boost (at the observer position) but it is not the same if we have two sources at the same position moving with different velocities (and observed by the same observer). We stress that this amplification can be very large for sources moving with quasi-relativistic velocities towards the observer (the Doppler factor gets very large in these cases). 
The object that transforms in a symmetric way under boosts of either the source or the observer is the specific intensity, i.e.\ the energy density per unit of solid angle, associated to a background or an extended source, since it does not depend on the distance to the source generating the signal. 

In addition to the change of amplitude, the source and observer velocity modify the polarization tensor through the effect of aberration. More precisely, we have shown that the polarization tensor is rotated with respect to the reference one. If we receive just one signal from the source (i.e.\ we do not have multiple images), then this rotation of the polarization tensor is fully degenerate with a reorientation of the source and therefore unobservable.

We then extended our derivation to the realistic case of a source emitting with an anisotropic pattern (such as a binary system of compact objects). In this case, we have shown that in full generality the amplitude of the wave emitted by a reference source and a boosted one (observed by the same observer) differ not only by the presence of a Doppler shift, but that there is an additional direction-dependent factor due to the fact that aberration affects the relative orientation of the emission direction with respect to the source intrinsic orientation. However, as for the polarization tensor, this effect is always degenerate with a reorientation of the source (that compensates for the effect of aberration of the emission direction), as was already shown in~\cite{Bonvin:2022mkw} at linear order in velocities. 

Based on the transformation laws that we derived, we then constructed a family of sources, with different peculiar velocities, that produce exactly the same observed signal. The existence of such a family of degenerate sources means that the source velocity \emph{cannot} be measured from the GW form, contrary to what was claimed in~\cite{PhysRevD.104.123025, PhysRevD.100.063012}. This statement holds for any velocity, even relativistic one. 


Finally we turned to a cosmological context, where reference source and observer can be naturally identified as comoving with the Hubble flow. We considered the case of a binary system of compact objects, and showed that, beside an unobservable spin phase of the helicity modes, resulting in an irreducible bias of the parameters describing the source orientation, peculiar velocities affect the wave amplitude. We stressed that, if the source redshift can be reconstructed via the observation of an electromagnetic counterpart, the strain is affected by both the source and the
observer velocity, while only the source velocity affects the amplitude in the absence of the redshift information. This is simply due to the fact that two different observers (one moving and one
at rest with respect to the Hubble flow) that are seeing the same source at the same redshift cannot
be both at the same luminosity distance from the source. Hence they do not measure the same
strain. 
A detailed study of how the source velocity affects the reconstruction of the source parameters  will be presented in~\cite{Pitrou2024}.

\subsection*{Acknowledgements}
We thank Jean-Philippe Uzan and Pierre Fleury for discussions during different stages of this work.  The work of Giulia Cusin is supported by CNRS and SNSF Ambizione grant \emph{Gravitational wave propagation in the clustered universe}. CB acknowledges funding from the European Research Council (ERC) under the European Union’s Horizon 2020 research and innovation program (Grant agreement No.~863929; project title ``Testing the law of gravity with novel large-scale structure observables"), and from the Swiss National Science Foundation. 

\newpage

\appendix

\section{Eikonal approximation in electromagnetism}\label{SecEikonalA}

In the case of electromagnetism, we consider an expansion for the vector potential $a_\mu$, analogous to~\eqref{ansatz}
\begin{equation}\label{ansatz2}
	a_{\mu}= \Re(A_{\mu}e^{i\omega \Phi})\,.
\end{equation}
With the Lorenz gauge condition $\nabla^\mu a_\mu = 0$, and following the same expansion in the geometric optics parameter $\omega$ of the Maxwell equations $\nabla^\alpha \nabla_\alpha a_\mu = R_\mu^{\,\,\nu}a_\nu$ as in section \ref{SecWKB} (see e.g. section 1.2 of \cite{Fleury:2015hgz}) we eventually get $k_\mu k^\mu=0$ and
\begin{equation}
2 k^\beta \nabla_\beta A_{\mu}+ A_{\mu} \nabla_\beta k^\beta=0\,.\label{go:linear2}
\end{equation}
Hence, with the same arguments as for gravitational waves we also obtain that $k^\mu \nabla_\mu k_{\nu}=0$, hence light is described by null geodesics, which are identified with light rays of the geometric optics description. The Lorenz gauge condition leads at leading order in $\omega$ to 
\begin{equation}
k^\mu A_{\mu}=0\,,
\end{equation}
which indicates that the polarization of the vector potential is a transverse vector. Separating $A_{\mu}$ into an amplitude and a polarization part as $ A_{\mu}=A  \epsilon_{\mu}$ with $A = \sqrt{A^\star_\mu A^\mu}$, $\epsilon_{\mu}\epsilon^{\mu}=1$ and $k^\mu \epsilon_\mu=0$\,, eq.~(\ref{go:linear2}) leads to two equations
\begin{subequations}
\begin{align}
& k^\mu \nabla_\mu A=-\frac{1}{2}\theta A; \quad \theta = \nabla_\mu k^\mu \label{conservationbis}\\
& k^\alpha \nabla_\alpha \epsilon_{\mu}=0 \label{conservation2bis}\,,
\end{align}
\end{subequations}
which are exactly like eqs.~\eqref{conservation} and \eqref{conservation2} with $H \to A$ and $\epsilon_{\mu\nu} \to \epsilon_\mu$. Hence, light polarization is parallel-propagated along the null vector $k^\mu$, and we obtain the covariant conservation of the flux (i.e.~$\nabla_\alpha (A^2k^\alpha)=0$), also interpreted as the conservation of photon number. Beyond geometric optics corrections become relevant at large but finite frequencies: the ray dynamics becomes affected by the evolution of the wave polarisation, hence rays can deviate from null geodesics, which is known as the gravitational spin Hall effect (see \cite{Oancea:2020khc,Frolov:2024ebe}  for recent studies). Furthermore, modifications to Maxwell theory of electromagnetism, such as a non-minimal coupling to curvature as in vector Horndeski theory~\cite{Horndeski:1976gi}, can also alter the previous equations by adding corrections~\cite{Lafrance:1994in}.

Following the same steps as for GW in the eikonal approximation, one deduces the scaling of the vector potential amplitude with propagation 
\be\label{ampl2}
A(\lambda)=A(\lambda_s)\frac{\chi(\lambda_s)}{\chi(\lambda)}\,,
\ee
analogous to eq.~\eqref{ampl}.

In Gaussian units, the energy momentum tensor of light is analogously obtained as 
\be\label{tmunuA}
t_{\mu\nu}=\frac{1}{8\pi}\left[E^2 + B^2\right]_{\rm av} = \frac{1}{8\pi}A^2 k_{\mu}k_{\nu}\,,
\ee
where $E$ and $B$ are the norms of the electric and magnetic fields associated to $a_\mu$. The associated energy density measured by an observer with velocity $u^\mu$ is
\be\label{rhoA}
\rho_{\rm light}= \frac{1}{8\pi c^2}A^2 \left(k_{\mu}u^{\mu}\right)^2\,,
\ee
hence the analogy between GW and light is obtained through the replacement $ c^2 H /\sqrt{8G} \to A$ in eqs.~\eqref{tmunuGW} and \eqref{rho} (note that $H$ is dimensionless but not $A$ in Gaussian units).

\section{Transformation of polarization vector under a boost}\label{Cyril}

We derive here the transformation properties of the polarization vector of photons under a boost.  For light, all polarization vectors $\tilde\epsilon_\mu = \epsilon_\mu + \xi k_\mu$ satisfy the gauge condition $\tilde\epsilon_\mu k^\mu = 0$. We usually select the unique $\xi$ such that it is also orthogonal to the observer (Coulomb gauge). For the boosted observer, the polarization chosen must satisfy $\tilde{\bm u} \cdot \tilde{\bm \epsilon} =0$, hence we deduce that $\xi = - \tilde{\bm u} \cdot {\bm \epsilon}/\tilde{\bm u} \cdot {\bm k}$ is needed. With \eqref{kdecomposition} this can be rephrased in a compact form as
\be
\tilde{\bm \epsilon}= {\bm \epsilon} + {\bm \epsilon}\cdot \tilde{\bm u}(\tilde{\bm u} - \tilde{\bm n}) = \tilde S({\bm \epsilon})\,,
\ee
where the screen projection operator is
\be\label{Eq:ScreenUse}
\tilde{S}_\mu^{\,\,\nu} = \delta_\mu^\nu + \tilde{u}_\mu \tilde{u}^\nu - \tilde{n}_\mu \tilde{n}^\nu\,,\qquad \tilde{\epsilon}_\mu = \tilde{S}_{\mu}^{\,\,\nu} \epsilon_\nu\,.
\ee
Note that we used the fact that ${\bm \epsilon}\cdot \tilde{\bm u} = {\bm \epsilon}\cdot \tilde{\bm n}$, ${\bm \epsilon}\cdot{\bm u} = {\bm \epsilon}\cdot {\bm n}$ since ${\bm \epsilon}\cdot{\bm k}=0$. These properties are then also satisfied for $\tilde{\bm \epsilon}$.  
The components are obtained by contraction with $\tilde{\bm e}_i $. We use first that $\tilde{\bm e}_i \cdot \tilde{\bm n} \equiv \widetilde{n_i}$ and $\tilde{\bm e}_i \cdot \tilde{\bm u} = 0$. Then to contract the first term on the rhs we use $\tilde{\bm e}_i = \Lambda_i^{\,\,j} {\bm e}_j + \Lambda_i^{\,\,0} {\bm u}$. Since ${\bm \epsilon} \cdot {\bm u} = 0$ this gives $\Lambda_i^{\,\,j} \epsilon_j$. Finally we use the identity $\gamma^2 v^2/(\gamma+1) = \gamma-1 $ to recast $\Lambda_i^{\,\,j}$. With the definitions $\epsilon_i \equiv {\bm \epsilon} \cdot {\bm e}_i$ and $\widetilde{\epsilon_i} \equiv \tilde{\bm \epsilon} \cdot \tilde{\bm e}_i$ (where the tilde stretches over both the vector name and the component to emphasize that we consider a different polarization vector that satisfies the new Coulomb gauge, and its components are also read in a different basis) we get
\be\label{Eq:epsilontilde1}
\widetilde{\epsilon_i} = \epsilon_i + (\gamma-1)({\bm{\hat v}} \cdot {\bm{\epsilon}}) \hat v_i  - \gamma ({\bm{v}} \cdot {\bm{\epsilon}}) \widetilde{n_i} =\left[\delta_{i}^j-\frac{1}{\mathcal{D}}\left((\gamma-1)\hat{v}_i + \gamma \beta n_i\right)\hat{v}^j\right]\epsilon_j\,, 
\ee
where in the last step we used \eqref{Eq:ntilde1}.

Let us show that the aberration formula \eqref{Eq:ntilde1} and the transformation of polarization \eqref{Eq:epsilontilde1} correspond to solid rotations of $n_i,\epsilon_i$ into $\widetilde{n_i},\widetilde{\epsilon_i}$ by constructing explicitly such rotations. For each wave direction $n_i$, the rotation is around an axis which is orthogonal to both $n_i$ and $\widetilde{n_i}$ (or orthogonal to both $n_i$ and $v_i$) and it is found from exponentiation of infinitesimal rotations as 
\be\label{JijtoRij}
R_{ij}(v_k,n_k) = \exp\left( \rotangle J_{ij} \right)\,,\qquad J_{ij} \equiv w_{i} n_j - w_j n_i\,,
\ee
with the unit direction of the transport (a unit vector tangential to the unit sphere) being
\be\label{Defvperpw}
w_i = \frac{\hat{v}^\perp_{i}}{\sqrt{1-({\bm {\hat{v}}}\cdot {\bm n})^2}}\,, \qquad \hat{v}^\perp_{i}\equiv \hat{v}_{i} - ({\bm {\hat{v}}}\cdot {\bm n}) n_i \,.
\ee 
The rotation angle $\rotangle$ in \eqref{JijtoRij} is found from (\ref{aberr}), and we obtain (with $0\le \rotangle \le \pi$)
\begin{subequations}
\begin{align}
\cos \rotangle &= n^i \widetilde{n_i} = \frac{1}{\mathcal{D}}(1+ \alpha {\bm {\hat{v}}}\cdot {\bm n}) = 1 - \frac{(\gamma-1)}{\mathcal{D}}[1-({\bm {\hat{v}}}\cdot {\bm n})^2]\,,\\
\sin \rotangle &= \frac{\alpha}{\mathcal{D}} \sqrt{1-({\bm {\hat{v}}}\cdot {\bm n})^2}\,,
\end{align}
\end{subequations}
where $\sin \rotangle$ was obtained from $\sqrt{1- \cos^2 \rotangle}$ and the identity \eqref{id}. It is clear that the rotation angle $\rotangle$ is direction dependent. In case the azimuthal direction of the spherical coordinates system is aligned with the velocity direction, it can be checked with the previous expressions that the aberration relation \eqref{EqAberrationCoordinates} is equivalent to $\tilde\theta = \theta - \rotangle$, as expected. The definition of the rotations~\eqref{JijtoRij} depends on $n_i$ and $v_i$, but to alleviate the notation we shall use $R_{ij} \equiv R_{ij}(v_k,n_k)$ when there is no ambiguity.

\begin{figure}[!htb]
\centering
\includegraphics[width=0.38\textwidth]{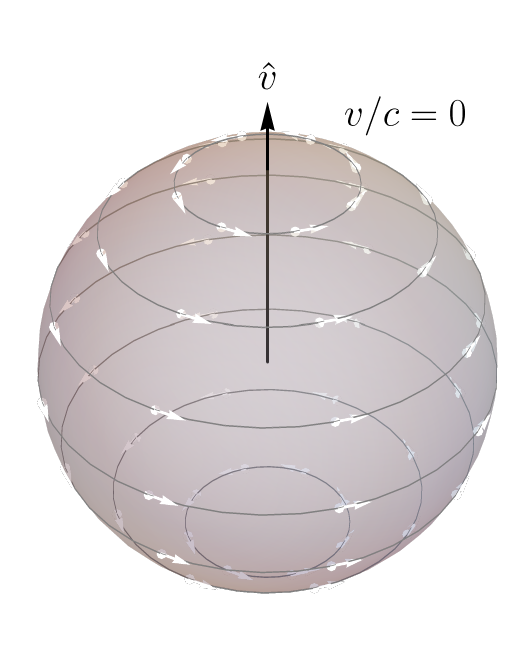}
\includegraphics[width=0.38\textwidth]{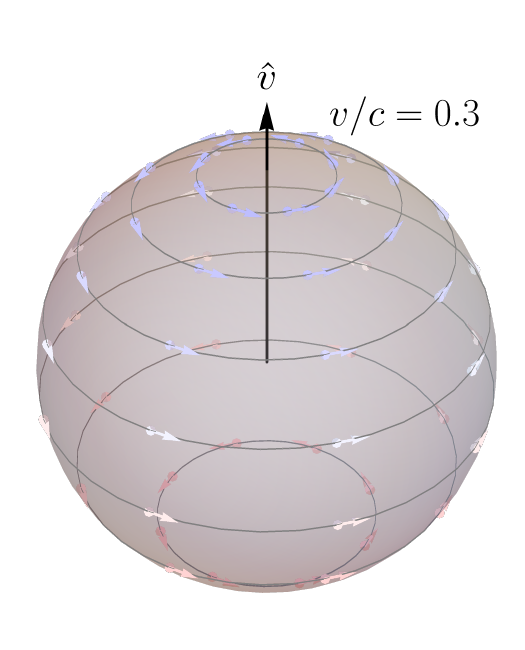}\\
\includegraphics[width=0.38\textwidth]{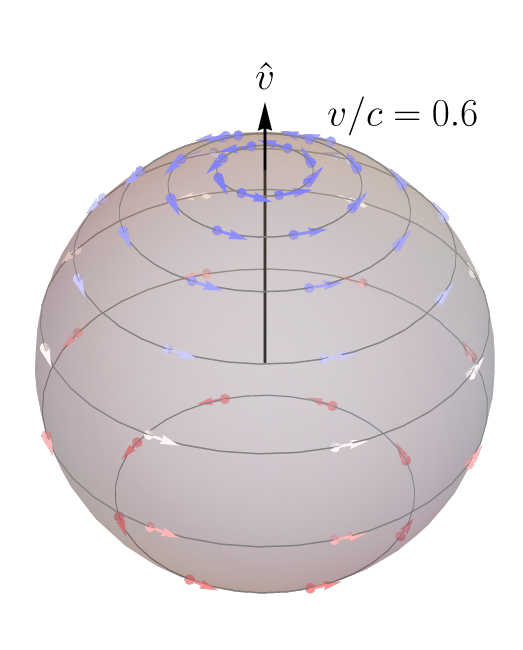}
\includegraphics[width=0.38\textwidth]{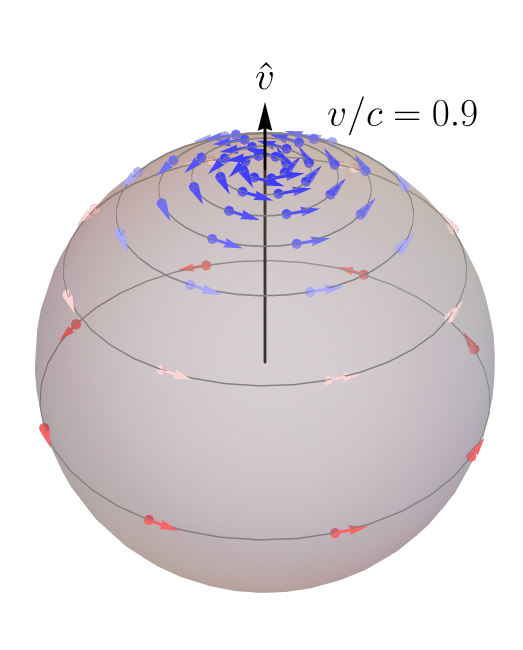}
\caption{\small {\it Top left:} A set of initial wave directions $n^i$ represented by points on the unit sphere along with the polarization vectors which for the purpose of illustration are chosen to be always aligned with ${\bm e}_\theta$. {\it Other panels:} Corresponding aberrated directions with the transformed polarizations, for several values of $\beta=v/c$ and when the velocity direction is aligned with the azimuthal direction of the spherical coordinate system. The blue hue (resp. red hue) is related to the amount of blueshifting, that is ${\cal D} > 1$ (resp. redshifting, that is ${\cal D}<1$). All directions are aberrated toward the velocity direction, however redshifting takes place when directions are in the lower hemisphere and satisfy $\cos \theta < (\sqrt{1-\beta^2}-1)/\beta$, see also bottom panels of Fig.~\ref{DopAngle}. With these adapted coordinates and the aberration relation \eqref{EqAberrationCoordinates}, it is clear that the great circles along which the rotations transport the directions (and parallel transport the polarizations) are the meridian circles, hence the aberrated polarization is always equal to ${\bm e}_\theta$ in this specific example.}
\label{FigAberration}
\end{figure}

Expliciting the exponentiation in \eqref{JijtoRij} and using that $w_i$ and $n_i$ are mutually orthogonal unit vectors, we get after resummation
\begin{align}\label{Rijresummed}
R_{ij} &= \delta_{ij} + (\cos \rotangle -1)(w_i w_j + n_i n_j ) + \sin \rotangle (w_i n_j - w_j n_i)\\
 &= \delta_{ij} + \frac{1-\gamma}{\mathcal{D}}\left[\hat{v}^\perp_i \hat{v}^\perp_j + n_i n_j \left(1-({\bm {\hat{v}}}\cdot {\bm n})^2\right)\right] + \frac{\alpha}{\mathcal{D}}\left(\hat{v}^\perp_i n_j - \hat{v}^\perp_j n_i \right)\,.\nonumber
\end{align}
It is immediate to check with the definition \eqref{Defvperpw}, the property $\hat{v}^\perp_i n^i = 0$ and the identity \eqref{id} that it is indeed an orthogonal matrix ($R_{ik} R_{i}^{\,\,k} = \delta_{ij}$).

It is now possible, using also that $n_i \epsilon^i=0$, to show that the expressions for the transformation of direction \eqref{Eq:ntilde1} and polarization \eqref{Eq:epsilontilde1} are simply 
\be\label{Rijne}
\widetilde{n_i} = R_i^{\,\,j} n_j\,, \qquad  \widetilde{\epsilon_i} = R_i^{\,\,j} \epsilon_j\,.
\ee
Hence, aberration acts as a local rotation on both the wave direction vector and the polarization vector. It transports the direction $n^i$ along the great circle generated by $w^i$, which is defined as the intersection of the unit sphere with the plane orthogonal to the axis of rotation $\epsilon^{ijk} n_j w_k$ (and containing the origin of coordinates). The transformation~\eqref{Rijne} also indicates that polarization is (parallel) transported along the same great circle. This is illustrated in Fig.~\ref{FigAberration}.

Finally, if we consider the components in the helicity basis $\epsilon_\pm ({\bm n})\equiv \epsilon_{i} {e}^i_{\mp}({\bm n})$ we get from the previous transformation of polarization and \eqref{TransfoHelicity}
\be
\tilde \epsilon_{\pm}(\tilde{{\bm n}}) \equiv \widetilde{\epsilon_{i}}  \tilde{e}^i_{\mp}(\tilde{{\bm n}}) = {\rm e}^{\pm {\rm i}\delta} \epsilon_{\pm}({\bm n})\,,
\ee
a transformation rule similar to \eqref{TransfoHelicityComponents1} but with the phase being $\pm\delta$ instead of $\pm 2 \delta$, reflecting the different spinorial nature of electromagnetic and gravitational waves.

\bibliography{GWBoostrefs}

\end{document}